\RequirePackage{fix-cm} 
\RequirePackage{amsmath}
\documentclass[twocolumn]{svjour3}

\smartqed  % flush right qed marks, e.g. at end of proof

\usepackage{times}

\usepackage[numbers]{natbib}

\usepackage{balance}
\usepackage[normalem]{ulem}

\usepackage{graphicx} % Required for inserting images
\usepackage{color}
\usepackage{amsmath}
\usepackage{pgfplots}
\usepackage{tikz}
\usetikzlibrary{shapes,arrows,positioning}
\usepackage{graphdb}
\pgfplotsset{compat=1.18} 
\usetikzlibrary{matrix}
\usepackage{booktabs}
\usepackage{ulem}
\usepackage{fullpage}

\makeatletter
\newenvironment{customlegend}[1][]{%
    \begingroup
    \pgfplots@init@cleared@structures
    \pgfplotsset{#1}%
}{%
    \pgfplots@createlegend
    \endgroup
}%

\def\addlegendimage{\pgfplots@addlegendimage}
\makeatother

\newcommand{\rev}{{\hat{~}}}
\newcommand{\U}{\mathcal{U}}

\newcommand{\dd}{\mathinner{.\,.}}
\newcommand{\rank}{\mathsf{rank}}

\newcommand{\dom}[1]{\ensuremath{\mathsf{dom}(#1)}}

\newcommand{\lab}[1]{\ensuremath{\mathsf{word}(#1)}}

\newcommand{\comp}[1]{\ensuremath{#1^{\leftrightarrow}}}

\newcommand{\adrian}[1]{{\color{brown!90!black}\textsc{Adrian:} #1}}

\newcommand{\no}[1]{}

\title{Evaluating Regular Path Queries \\ on Compressed Adjacency Matrices\thanks{\tiny This work was supported by ANID – Millennium Science Initiative Program – Code ICN17\_002, and Fondecyt Grant 1-230755, Fondecyt Grant 1221926; 
CITIC is funded by Xunta de Galicia and CIGUS; GAIN/Xunta de Galicia Grant ED431C 2021/53 (GRC); Xunta de Galicia/FEDER-UE Grant IN852D 2021/3; MCIN/AEI and NextGenerationEU/PRTR Grants [PID2020-114635RB-I00, TED2021-129245B-C21]. A preliminary version of this paper appears in {\em Proc. SPIRE 2023}.
}}

\author{Diego Arroyuelo \and Adrián Gómez-Brandón \and Gonzalo Navarro}

\institute{D.\ Arroyuelo \at
              IMFD \& DCC, Escuela de Ingenier\'ia, Pontificia Universidad Cat\'olica de Chile, Santiago, Chile\\ 
              \email{diego.arroyuelo@uc.cl}
           \and
            A.\ G\'omez-Brand\'on \at
              IMFD \& CITIC, Universidade da Coru\~na, A Coru\~na, Spain\\
              \email{adrian.gbrandon@udc.es}
           \and 
           \Letter \hspace{0.1cm} G.\ Navarro \at
              IMFD \& DCC, University of Chile, Santiago, Chile\\
              \email{gnavarro@dcc.uchile.cl}            
}

\begin{document}

\maketitle

\begin{abstract}
Regular Path Queries (RPQs), which are essentially regular expressions to be matched against the labels of paths in labeled graphs, are at the core of graph database query languages like SPARQL. A way to solve RPQs is to translate them into a sequence of operations on the adjacency matrices of each label. We design and implement a Boolean algebra on sparse matrix representations and, as an application, use them to handle RPQs. Our baseline representation uses the same space as the previously most compact index for RPQs and outperforms it on the hardest types of queries---those where both RPQ endpoints are unspecified. Our more succinct structure, based on $k^2$-trees, is 4 times smaller than any existing representation that handles RPQs, and still solves complex RPQs in a few seconds. Our new sparse-matrix-based representations dominate a good portion of the space/time tradeoff map, being outperformed only by representations that use much more space. They are also of independent interest beyond solving RPQs.
\end{abstract}

\section{Introduction and Related Work}

Graph databases have emerged as a crucial tool in several applications such as web and social network analysis, the semantic web, and modeling knowledge, among others. We are interested in labeled graph databases, where the graph edges have labels. One particular way of querying graph databases is by means of basic graph patterns (BGPs, for short), which are small subgraphs with constant or variable nodes and edge labels that are to be matched homomorphically in the graph database. 
%In our example, the BGP $\{ \mathtt{?x} \xrightarrow{\mathsf{walk}} \mathtt{?y},~ \mathtt{?y} \xrightarrow{\mathsf{walk}} \mathtt{?z},~ \mathtt{?z} \xrightarrow{\mathsf{walk}} \mathtt{?x} \}$ looks for three places within walking distance to each other.
BGPs are strongly related to relational database joins \cite{HoganRRS19}. Another important kind of queries that are more exclusive of graph databases are the \emph{regular path queries} (RPQs, for short), which search for paths of arbitrary length matching a regular expression on their edge labels \cite{AnglesABHRV17}. For example, in the simple RDF model \cite{rdf}, one can represent points of interest in New York City as nodes in a graph, and have edges such as $x \xrightarrow{\mathsf{walk}} y$ indicating that $x$ is within a short walking distance of $y$, as well as edges of the form $x \xrightarrow{\mathsf{L}} y$ if subway stations $x$ and $y$ are connected directly by subway line $\mathsf{L}$. Then the RPQ
%`$\text{Central Park}~\mathsf{walk}/(\mathsf{N} | \mathsf{Q} | \mathsf{R})^+/\mathsf{walk}~\mathtt{?y}$',
`$\textsf{Central Park}~\mathsf{walk}/(\mathsf{O} | \mathsf{R})^+/\mathsf{walk}~\mathtt{?y}$',
asks for all sites $\mathtt{?y}$ of interest that are reachable from Central Park by using subway lines \textsf{O}ne  or \textsf{R}, through one or more stations and allowing a short walk before and after using the subway.

%Mention that RPQs appear in various query languages such as SPARQL, GQL (?), ...  \cite{GQL}
RPQs are at the core of current graph database query languages, extending their expressiveness. In particular, the SPARQL 1.1 standard includes the support for \emph{property paths}, that is, RPQs extended with inverse paths (known as two-way RPQs, or 2RPQs for short) and negated label sets. As SPARQL has been adopted by several systems, RPQs have become a popular feature~\cite{AnglesABHRV17}: out of 208 million SPARQL queries in the public logs from the Wikidata Query Service~\cite{MalyshevKGGB18}, about 24\% use at least one RPQ feature~\cite{BonifatiMT19}. Further developments like PGQL~\cite{RestHKMC16}, Cypher~\cite{FrancisGGLLMPRS18}, G-CORE~\cite{AnglesABBFGLPPS18}, TigerGraph~\cite{DeutschXWL20}, and GQL~\cite{GQL}, to name some of the most popular ones, also support RPQ-like features. 
%So, efficiently managing RPQs is crucial for the systems supporting these languages. 

%Several approaches to handle RPQs, esp. the product graph.
Handling (2)RPQs can be computationally expensive as they usually involve a large number of paths \cite{MartensNPRVV23}, mostly for regular expressions using Kleene stars. There are two main algorithmic approaches to support them \cite{YakovetsGG16}: (1) to represent the regular expression of the 2RPQ using a finite automaton, which is then used to search over the so-called product between the automaton and the database graph~\cite{MendelzonW95}; and (2) to extend the relational algebra to support computing the transitive closure of binary relations in order to evaluate regular expressions having Kleene stars \cite{LMpods12}. Although most theoretical results on 2RPQs have followed the first approach, property path evaluation in SPARQL has followed the second one \cite{YakovetsGG16}.

%Recent approaches to compressed data structures, our ICDE paper.

Recent research introduced not only time- but also space-efficient solutions for evaluating graph joins~\cite{ANRRtods22,AHNRRS21,BCdBFNsupe22,AGHNRRStods24}. With the big graphs available today, this is an important step towards in-memory processing of graph queries. In particular, the Ring data structure~\cite{AHNRRS21,AGHNRRStods24} is able to represent a labeled graph in space close to its plain representation, while supporting worst-case optimal joins (used, as we said, for BGP queries).  Moreover, by using little extra space the Ring can be used to support 2RPQs efficiently \cite{AHNRicde22,AGHNRvldbj24}, using the product-graph approach~\cite{MendelzonW95}.

%Alternative approach based on matrix operations: Boolean sum, Boolean multiplication, transitive closure.  Find a reference on this, I guess it exists... \diego{Losemann y Martens basan la evaluación en algebra de matrices, excepto para la concatenación que la hacen con un join. Sin embargo, es un paper teorico sobre la complejidad de evaluar RPQs, no dice mucho de la implementacion.}
%Typically disregarded because matrices are huge, but otoh they tend to be sparse.
%We explore this approach: translate RPQs into operations on sparse matrices. Those are represented with the well-known k2tree. We show how to do this translation and handle the particularities of RPQs. The result is the most succinct graph db representation ( 5.x bpt in our wikidata experiments) while being competitive with approaches that use much more space. 

\paragraph{Our contribution.}
In this paper, we introduce a space-efficient approach for evaluating 2RPQs that, essentially, represents the subgraph corresponding to each graph label $p$ using a sparse representation of its Boolean adjacency matrix $M_p$. We evaluate 2RPQs by translating them into classic operations on Boolean matrices \cite{LMpods12}.
%(such as, e.g., sums, products, and transitive closures). 
This approach is typically disregarded because matrix sizes are quadratic on the number of graph nodes, but we exploit the sparsity of those matrices to represent them efficiently with two approaches:
\begin{enumerate}
\item \label{it:k2trees} We use $k^2$-trees \cite{BLNis13} to represent each RDF predicate in compressed form. Although $k^2$-trees have been already used to handle triple matching and binary joins \cite{AGBFMPN13} and full BGPs \cite{ANRRtods22}, their use for supporting 2RPQs is new and requires novel algorithms. We show how to translate 2RPQs into matrix operations, particularly to Boolean sums, multiplications, and transitive closures, among other particularities of 2RPQs. We improve and extend known algorithms for the Boolean sum on $k^2$-trees \cite{QFPLG19}, and develop new ones for sparse matrix multiplication (following a quite natural recursive strategy) and transitive closure (a not so obvious strategy we develop that yields the same time complexity of a matrix multiplication).

\item \label{it:baseline} We also adapt and implement an uncompressed baseline for sparse Boolean matrices based on the CSR and CSC formats \cite[Sec.~3.4]{Saa03}. Our baseline implements state-of-the-art algorithms for sparse matrices, like Shoor's multiplication \cite{Schoor82} adapted to the Boolean case, and an algorithm based on finding strongly connected components \cite{Purdom70,Tar72} for the transitive closure. We implement those algorithms with special care on minimizing the working space. 
\end{enumerate}

Our main results can be summarized as follows:
\begin{itemize}
\item Our $k^2$-tree based representation (item \ref{it:k2trees} above) is the most space-efficient graph database representation so far. It uses nearly 4 bytes per graph edge on a Wikidata graph, which is 4 times less than the previously most compact representation---the Ring \cite{AHNRicde22,AGHNRvldbj24}---, 6.5 times less than Ring$_\text{AB}$---a larger and faster Ring variant---, and 14--22 times smaller than classical systems. In exchange, our structure is about 3 times slower than the Ring and 8 times slower than Ring$_\text{AB}$, though it still solves most 2RPQs within a few seconds. On the harder queries, however---those featuring both variable extremes---, our structure is slightly faster than the Ring and about 3 times slower than Ring$_\text{AB}$.
\item Our baseline (item \ref{it:baseline} above)  uses about 4 times the space of our $k^2$-tree based structure and is considerably faster. Its space matches that of the Ring, and outperforms it by a factor of 2.3 on the harder 2RPQs, still being 1.5 times slower than Ring$_\text{AB}$.
\end{itemize}

In general, our matrix-based implementations dominate the space/time tradeoff map of structures solving RPQs, yielding only to the Ring (which uses 4 times more space than $k^2$-trees, and stands out only on the easy queries) and to Ring$_\text{AB}$ (which uses 6.5 times more space than $k^2$-trees). A byproduct of our work yields a third relevant contribution:
\begin{itemize}
\item 
Both sparse Boolean matrix algebra implementa\-tions---from items \ref{it:k2trees} and \ref{it:baseline} above---are of independent interest, and can be used in many other applications where operations like Boolean sums (and others like conjunction, difference, exclusive-or, etc.), multiplications, and transitive closures are of use. We leave public implementations of both.
\end{itemize}

Compared to an early conference version of this paper \cite{AGNspire23}, the present article includes improved algorithms  for sum-like operations, multiplications, and especially transitive closures, on both the baseline and the $k^2$-tree based representations, a multithreadad implementation of the $k^2$-tree based algorithms, complete time complexity analyses of all the algorithms, and improved and extended experimental results.

\section{Basic Concepts}

\subsection{Labeled Graphs and Regular Path Queries (RPQs)} 
\label{sec:labeled-graphs-rpqs}

Let $\U$ be a totally ordered, countably infinite set of {\em symbols} or {\em constants}, which we call the \emph{universe}. A \emph{directed edge-labeled graph} $G \subseteq \U^3$ is a finite set of triples $(s, p, o) \in \U^3$ encoding the graph edges $s \xrightarrow{p} o$ from vertex $s$ to vertex $o$ with edge label $p$. In the RDF model~\cite{rdf} (which has gained popularity in representing directed edge-labeled graphs), $s$ is called a \emph{subject}, $p$ a \emph{predicate}, and $o$ an \emph{object}. 

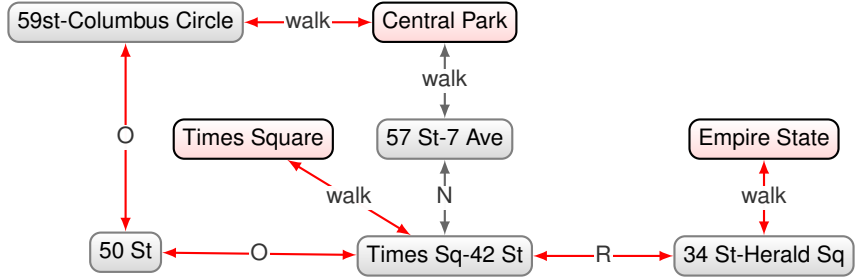
\begin{figure*}[tb]
\setlength{\hgap}{4.2cm}
\setlength{\vgap}{1cm}
\centering
\begin{tikzpicture}

\newcommand{\bend}{10}
\newcommand{\bbend}{25}
\newcommand{\colorntgt}{red!15}

\node[iri,anchor=mid, draw=black, top color=white, 
     bottom color=\colorntgt] (cp) {Central Park};

\node[iri,anchor=mid, xshift=-\hgap] (59-s) {59st-Columbus Circle}
 edge[arroutin, draw=red] node[lab] {walk} (cp);

\node[iri, anchor=mid,below=2.5\vgap of 59-s] (50-s) {50 St}
 edge[arroutin, draw=red] node[lab] {O} (59-s);
 
\node[iri, anchor=mid,below=\vgap of cp] (57-s) {57 St-7 Ave}
 edge[arroutin] node[lab] {walk} (cp);
 
\node[iri, anchor=mid,below=\vgap of 57-s] (42-s) {Times Sq-42 St}
 edge[arroutin] node[lab] {N} (57-s)
 edge[arroutin, draw=red] node[lab] {O} (50-s);

\node[iri, anchor=mid,below=\vgap of cp,xshift=-0.6\hgap, draw=black, top color=white, bottom color=\colorntgt] (ts) {Times Square}
 edge[arroutin, draw=red] node[lab] {walk} (42-s);

\node[iri, anchor=mid,below=\vgap of cp,xshift=\hgap, draw=black, top color=white, bottom color=\colorntgt] (es) {Empire State};

\node[iri, anchor=mid,below=\vgap of es] (34-s) {34 St-Herald Sq}
edge[arroutin, draw=red] node[lab] {walk} (es)
edge[arroutin, draw=red] node[lab] {R} (42-s);

\end{tikzpicture}
\caption{Graph with points of interest and subway line connections across New York City. Red edges represent the paths that hold RPQ `$\text{Central Park}~\mathsf{walk}/(\mathsf{O} | \mathsf{R})^+/\mathsf{walk}~\mathtt{?y}$' and the red nodes are its solutions.  \label{fig:ny}}
\end{figure*}

For a graph $G$, we define its set of edge labels as $P = \{p~|~\exists~s, o, (s,p,o)\in G\}$. Similarly, let $V = \{ x~|~\exists\, y,z,~(x,y,z) \in G \vee (z,y,x)\in G\}$ be the set of graph nodes. We assume that the graph nodes  have been mapped to integers in the range $[1\dd |V|]$.
A path $\rho$ from a node $x_0$ to node $x_n$ in a graph $G$ is a string $x_0p_1x_1\cdots x_{n-1} p_n x_n$ such that $(x_{i-1}, p_i, x_{i}) \in G$ for $1 \le i \le n$. Given a path $\rho$, we denote $\lab{\rho} = p_1\cdots p_n$ the string labeling path $\rho$. Two-way RPQs (2RPQs) also allow traversing reversed edges. Hence, we define the set of inverse labels as $\rev{P} = \{\rev{p}~|~p \in P \}$, and $\comp{P} = P \cup \rev{P}$ the set of predicates and their inverses. We define the {\em inverse graph} as $\rev{G} = \{(y, \rev{p}, x) ~|~ (x,p,y)\in G\}$, and its {\em completion} as $\comp{G} = G\cup\rev{G}$.
A \emph{two-way regular expression} (2RE) is then formed from the rules:
\begin{enumerate}
\item $\varepsilon$ is a 2RE.
\item If $c \in \comp{P}$, then $c$ is a 2RE.
\item If $E$, $E_1$ and $E_2$ are 2REs, so are $E^*$ (Kleene star), $E_1/E_2$ (concatenation), and $E_1~|~E_2$ (disjunction).  
\end{enumerate}

We also abbreviate $E^*/E$ as $E^+$ and $\varepsilon | E$ as $E^{?}$. 

The {\em language} $L(E)$ of $E$ is defined exactly as that of the regular expressions over the alphabet $\comp{P}$ of terminals, and we say that a path $\rho$ \emph{matches} a 2RE $E$ iff $\lab{\rho} \in L(E)$.

Let $\phi$ denote a set of variables, $\mu:\phi \rightarrow \U$ denote a partial mapping from variables to constants in $\U$, and $\dom{\mu}$ denote the set of variables for which $\mu$ is defined. If $E$ is a 2RE, $s\in \phi\cup\U$ and $o\in \phi\cup\U$, then $(s, E, o)$ is a {\em two-way regular path query}, or 2RPQ for short.
Let $x_\mu$ be defined as $\mu(x)$ if $x\in\dom{\mu}$, or $x$ otherwise. We define the {\em evaluation} of $(s, E, o)$ on $\comp{G}$ as:
\begin{align*}
&(s, E, o)(\comp{G}) = \{\mu~|~ \dom{\mu} = \{s, o\}~\cap~\phi \text{ and} \\
&\text{there exists a path } \rho \text{ from }
 s_\mu \text{ to } o_\mu \text{ in } \comp{G} \text{ matching } E\}.
\end{align*}

In other words, the result of evaluating a 2RPQ $(s, E, o)$ on $\comp{G}$ is the set of all pairs of constants $(s_\mu, o_\mu)$ for which there exists a path $\rho = s_\mu p_1\cdots p_n o_\mu$ in $\comp{G}$ such that $\lab{\rho} \in L(E)$.
Figure~\ref{fig:ny} illustrates our example RPQ on a small graph; the result of its evaluation is $\{ (\mathsf{Central~Park,Times~Square}), (\mathsf{Central~Park,Central}$ $\mathsf{Park}), (\mathsf{Central Park,Empire State})\}$.

\subsection{An Algebra on Boolean Matrices} \label{sec:matrix-algebra}

%Define the Boolean operations $A+B$, $A\times B$, $A^*$, $A^+$, and some extensions we will need: $A^T$, $[row]A$, and $A[col]$. The last two return a matrix of the same dimensions of $A$ but with only that row or col, the rest are zeros.

Let $A=(a_{i,j})_{1\le i, j\le n}$ and $B = (b_{i,j})_{1\le i, j\le n}$ be square $n\times n$ Boolean matrices. We define the following operations of interest for our work: %\gonzalo{Mencionar la notación $a_{i,j}$ para $A$}
\begin{itemize}
    \item \textbf{Transpose}: $A^T$, where $a^T_{i,j} = a_{j,i}$.
    \item \textbf{Sum}: $A+B = C=(c_{i,j})$, where $c_{i,j} = a_{i,j} \vee b_{i,j}$.
    \item Other sum-like operations like $A \cap B$ (where $\vee$ above is replaced by $\land$), $A - B$ (where $\vee$ is replaced by $\land \neg$), and $A\oplus B $ (where $\vee$ is replaced by exclusive-or).
    \item \textbf{Product}: $A\times B = C$, for $c_{i,j} = \bigvee_{1\le k\le n}{a_{i,k} \wedge b_{k,j}}.$
    \item \textbf{Exponentiation}: $A^k = \prod_{i=1}^{k}{A}$, that is, $A \times \cdots \times A$, writing $A$ $k$ times. % (A^{k/2})^2$, which takes $\log{k}$ multiplications.
    \item \textbf{Transitive closure}: $A^+ = A+A^2+\cdots + A^n$.
    \item \textbf{Reflexive-transitive closure}: $A^* = I + A^+$, where $I$ is the identity matrix.
    \item \textbf{Row restriction}: $\langle r\rangle A$, a matrix whose row $r$ equals row $r$ of $A$, the remaining cells are $0$.
    \item \textbf{Column restriction}: $A\langle c\rangle$, a matrix whose column $c$ equals column $c$ of $A$, the remaining cells are $0$. 
    \item \textbf{Cell restriction}: $\langle r\rangle A\langle c\rangle$, a matrix whose cell $(r,c)$ equals entry $A[r][c]$; the other cells are $0$.
\end{itemize}

The implementation of these operations on sparse matrix representations is relatively straightforward, except for the multiplication and transitive closures. We review those algorithms next.

\subsection{Boolean Matrix Multiplication and Transitive Closure} \label{sec:clausura}

The multiplication of two $n\times n$ Boolean matrices $A$ and $B$, of $a$ and $b$ non-zero entries, respectively, is one of the most important operations of the Boolean-matrix algebra, because of its applications in context-free parsing \cite{Valiant75}, context-free path queries on labeled graphs \cite{AEGgrades21}, triangle detection in graphs \cite{IRsicomp78,Yu18}, and on computing the transitive closure of Boolean matrices \cite{FMswat71,Munro71,Furman70}. To illustrate its importance in the context of directed graphs, if $A$ is a Boolean matrix representing the adjacency matrix of the graph, then $A^2 = A\times A$ is such that $A^2[i][j] = 1$ iff there is a path of length exactly 2 between nodes $i$ and $j$. Also, by computing $(I + A)^2$ one obtains the Boolean matrix indicating the pairs of nodes $(i,j)$ such that there is a path of length at most 2 between them. This can be generalized to any positive $k$-th power \cite{Yannakakis90}.

The most efficient algorithms for matrix multiplication work on algebraic rings, whereas $(0, 1, \vee, \wedge)$, the Boolean case, is just a semiring as there is no additive inverse. For instance, Strassen’s algorithm \cite{Stra69} needs subtraction. A natural solution for the Boolean case is, however, to take the two values as integers, to then apply some fast multiplication algorithm. The result is then translated back to a Boolean matrix by replacing any non-zero value by a 1, whereas 0s remain unchanged. The fastest known matrix multiplication algorithm, by Coppersmith and Winograd, runs in time $O(n^{2.373})$ %2.3728639})$ 
\cite{CWjsc90,Williams12}. Very recent advances \cite{Faw22} suggest that this exponent can be further pushed towards the lower bound $\Omega(n^2)$.
%It is unknown whether the matrix multiplication exponent $\omega$ can be improved to $2 \le \omega < 2.3728639$. 
Another approach is that of \emph{combinatorial algorithms}, which use combinatorial properties of Boolean matrices to improve computation time. A typical example of this line is the (original) Four-Russians approach by Arlazarov et al.~\cite{ADKF1970}, which runs in time $O(n^3/\log^2{n})$ on a word RAM of $\Theta(\log n)$ bits \cite{Yu18}. After several progressive  improvements, Yu \cite{Yu18} introduced an algorithm that runs in time $O(n^3\text{poly}(\log\log{n})/\log^4{n})$.

For sparse matrices,  Yuster and Zwick \cite{YZtalg05} introduce an algorithm that carries out $O(m^{0.7}n^{1.2} + n^{2+o(1)})$ algebraic operations, where $m = \max(a, b)$. As noticed by Yuster and Zwick, their algorithm runs in almost optimal $O(n^{2+o(1)})$ time when $m \le n^{1.14}$, and it outperforms Coppersmith and Winograd’s algorithm when $m \le n^{1.68}$. These algorithms are impractical in general because of big constants hidden in the asymptotic notation. A more practical one, by Amossen and Pagh \cite{AP09}, has output-sensitive time complexity $O(n^{2/3}z^{2/3}+n^{0.862}z^{0.408})$, where $z$ is the number of 1s in the output matrix.
In our baseline, we implement the algorithm of Schoor \cite{Schoor82}, which seems to be the most practical one. It takes $O(a b/n)$ time on average if the 1s are uniformly distributed, using $O(a+b)$ space to represent the matrices. It intersects the nonempty columns of $A$ with the nonempty rows of $B$, and adds to the result the Cartesian product of all the cells in the matching columns and rows. 

\medskip

Regarding the transitive closure $A^+$ of a Boolean matrix $A$ (again, with $a$ non-zero entries), a classic result by Warshall \cite{War62} achieves $O(n^3)$ time, just like a naive matrix multiplication. 
%It does not take advantage, however, of the case $a\ll n^2$ 1s. 
%\diego{El problema es equivalente a multiplicar matrices, entonces cualquier algoritmo necesita tiempo al menos $n^2$, quizas solo el espacio cuadratico sea problema.}. Also, it uses $n^2$ bits of space. 
Although $A^+ = A + A^2 + \cdots + A^n$, Furman \cite{Furman70} showed that only $O(\log{n})$ steps of the following process are needed. First, define $A_1 = A$, and then $A_{2k} = A_{k} + A_{k}^2$, for $k=1,2,\ldots,\lceil\log_2 n\rceil$. By embedding the Boolean matrix into a ring, one can then achieve time $O(n^{\omega}\log{n})$. Munro \cite{Munro71} and Fischer and Meyer \cite{FMswat71} showed that Boolean matrix multiplication and transitive closure have essentially the same complexity, meaning that only one matrix multiplication is enough to compute the transitive closure. Hence, all running times we gave for matrix multiplication are valid for transitive closure. 

A key idea for sparse matrices, which we implement in our baseline, is to detect the strongly connected components (scc) of the graph represented by the matrix, which can be done in $O(a)$ time \cite{AHU83,Sha81,Tar72,Dij76}. Every node can reach every other within each component, and the graph of the components (where we collapse all the vertices of each component into one) is acyclic, so reachability is easily computed on it.
Purdom \cite{Purdom70} introduced such an algorithm based on computing the scc, which runs in $O(a + \mu n)$ time, where $\mu$ is the number of scc. Munro’s algorithm \cite{Munro71} also computes the scc, yet it uses matrix multiplication to compute the transitive closure on the scc adjacency matrix. Nuutila \cite{Nuutila94} introduces an improved algorithm based on the same approach, which has good practical performance. Penn \cite{Penn06} introduces a sparse-matrix representation called Zero-Counting by Quadrants (ZCQ) and then shows how to use it to carry out matrix multiplication to compute the transitive closure, as in Munro’s algorithm. The approach is shown to be competitive in practice \cite{Penn06}. The particular matrix multiplication algorithm used by Penn mimics the one in Eq.~(\ref{eq:matrix-multiplication}), and we use it as inspiration to develop a novel transitive closure algorithm on $k^2$-trees.

Regarding its application to database management systems, several practical ideas have been proposed, such as the least-fixed point approach by Aho and Ullman \cite{AUpopl79} (and further improvements, see the excellent description by Nuutila \cite[Ch.~2]{Nuutila95}), graph traversals \cite{Yannakakis90,TKLgrades19}, and hybrid approaches \cite{Jakobsson91} mixing several of the above approaches. Amossen and Pagh \cite{AP09} use Boolean matrix multiplication to efficiently handle join-project queries, outperforming classical approaches in most cases.

\subsection{$K^2$-trees} \label{sec:k2trees}

%As a tool to represent sparse Boolean matrices. For $k=2$. Implemented over LOUDS-like. Explain the z-order, 0123 for top-left, top-right, bottom-left, bottom-right. Give the name signature to the 4 bits that describe each node.

\begin{figure*}[t]
    \begin{minipage}{0.3\textwidth}
    \centering
    \includegraphics[width=0.5\textwidth]{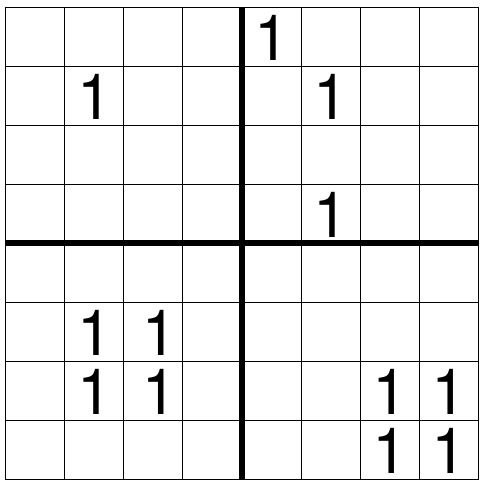}
    \end{minipage}
    %\hspace{1.5cm}
    \begin{minipage}{0.65\textwidth}
    \centering
    \includegraphics[width=\textwidth]{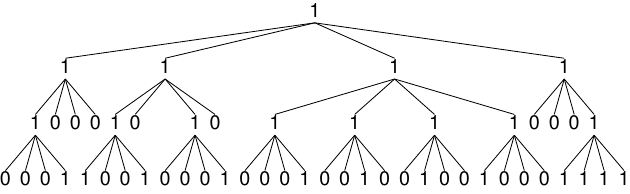}
    \end{minipage}
    
    \vspace{0.65cm}
    \begin{minipage}{0.3\textwidth}
       ~ 
    \end{minipage}
    \begin{minipage}{0.65\textwidth}
    \centering
    \includegraphics[width=\textwidth]{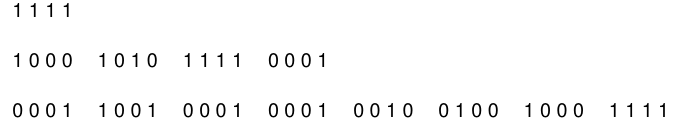}
    \end{minipage}
    \caption{A Boolean matrix (above, left), its corresponding $k^2$-tree representation (above, right), and the corresponding level-wise bitvector representation of the tree (below).}
    \label{fig:k2tree}
\end{figure*}

A $k^2$-tree \cite{BLNis13} is a data structure able to space-efficiently represent binary relations, point grids, and graphs. We will use it in this paper with $k=2$ to represent Boolean matrices, as follows. Let $A$ be a $v \times v$ Boolean matrix, assuming $v$ is a power of $k=2$.\footnote{If $v$ is not a power of 2 we round it up to the next power, leaving the extended cells empty. This imposes almost no extra overhead on the $k^2$-tree representation.} The root node of the $k^2$-tree represents the whole matrix $A$. Then, $A$ is divided into 4 equally-sized quadrants, $A = {A_0 ~ A_1 \choose A_2 ~ A_3}$, such that submatrix $A_0$ is represented recursively by the first child of the root, $A_1$ by the second child, and so on. The process stops as soon as one gets into an empty submatrix, which is represented by a leaf node, or else when the submatrix is a single cell. Each node in this tree has $k^2=4$ children.
Figure \ref{fig:k2tree} shows the $k^2$-tree representation of a sample Boolean matrix. Nodes representing a non-empty submatrix are marked with a \textsf{1}, otherwise the mark is a \textsf{0}.
This order in which quadrants are represented (i.e., top-left, top-right, bottom-left, and bottom-right) is known as z-order. The resulting tree height is $\log_4 v^2 = \log_2 v$, and the leaves list the 1s of $A$ in a left-to-right order imposed by the z-order. Concretely, the positions $A[i,j]=1$ are listed by increasing value of $\textsf{z-order}(i,j)$, which is computed as follows: since $i$ and $j$ are integers of $\log_2{v}$ bits each, $\textsf{z-order}(i,j)$ is the $(2\log_2{v})$-bit integer number obtained by interleaving the bits of the binary encodings of $i$ and $j$.

To represent this tree space-efficiently, we traverse it in level order. At each node, we write its 4-bit signature (which represents the node) indicating whether each of the 4 children represents an empty submatrix or not. For instance, the signature \texttt{0110} indicates that quadrants $0$ and $3$ of the submatrix represented by the current node are empty, whereas $A_1$ and $A_2$ (second and third children) are non-empty. The result is a bitvector $L[1\dd 4n]$, where $n$ is the number of internal nodes in the tree. Each tree node is represented by the position of the first bit of its signature. Given a node $i$, its $j$-th child ($1\le j\le 4$) is represented at position $4\cdot\rank_1(L, i) + 1$, where $\rank(L, i)$ counts the number of 1s in $L[1\dd i]$ in $O(1)$ time using $o(n)$ additional bits of space \cite{Cla96,Mun96}. Figure \ref{fig:k2tree} (below) shows the bitvector representation of the $k^2$-tree representing a Boolean matrix. For clarity, the three levels of bitvector $L$ are shown separately; the actual representation is the concatenation of these bitvectors.

The $k^2$-tree representation is especially useful for sparse matrices. Let matrix $A$ have $a$ 1s. Then, in the worst case every 1 induces a node (i.e., a 4-bit signature) in every level of the $k^2$-tree, for a total of $4a\log_2 v$ bits. Not all those induced nodes can be different, however: in the worst case all the
$k^2$-tree nodes up to level $\lfloor \log_4 a \rfloor$ exist, and from there on each 1 of $A$ has its own path; this adds up to $4a\log_4(v^2/a)+4a/3+O(1)$ bits. The figures further improve when the 1s are clustered in $A$ \cite{dBGLNSjcss22}.

We note that constant-time $\rank$ is possible in the so-called transdichotomous RAM model of computation, where we assume that the computer word holds $\Theta(\log v)$ words in order to represent $\log_2 v$-bit coordinates in $O(1)$ words (and thus handle them in $O(1)$ time). We assume this computation model as well.

\section{Evaluating RPQs using Boolean Matrix Algebra} \label{sec:evaluating-rpqs}

For a given directed edge-labeled graph $G$ of $n$ edges, let $P$ be the corresponding set of graph labels as defined in Section \ref{sec:labeled-graphs-rpqs}. In our approach, for every $p \in P$ we define a $|V|\times |V|$ Boolean matrix $M_p$, such that $M_p[x][y] = 1$ iff $(x, p, y) \in G$. We translate an RPQ into operations on those matrices, so that the resulting Boolean matrix contains all pairs $(x, y)$ that match the regular expression. 
We define next the recursive formulas ${\cal M}$ to translate 2RPQs into matrix operations, following Losemann and Martens' work \cite{LMpods12}. We start with the base cases: 
\begin{itemize}
    \item ${\cal M}(\varepsilon) = I$, the identity matrix.
    \item ${\cal M}(p) = M_{p}$, for $p \in P$. 
    \item ${\cal M}(\rev p) = M_{p}^T$, for $p \in P$.
\end{itemize}
Next, let $E_1$ and $E_2$ be 2RPQs. We define the following recursive rules: 
\begin{itemize}
    \item ${\cal M}(E_1 ~|~ E_2) = {\cal M}(E_1) + {\cal M}(E_2)$
    \item ${\cal M}(E_1 / E_2) = {\cal M}(E_1) \times {\cal M}(E_2)$
    \item ${\cal M}(E_1^+) = {\cal M}(E_1)^+$
    \item ${\cal M}(E_1^*) = I + {\cal M}(E_1)^+$, where $I$ is the corresponding identity matrix.
\end{itemize}
Then, given a 2RPQ $R = (x,E,y)$, we extend $\cal M$ to evaluate it as follows:
\begin{enumerate}
    \item If $x$ and $y$ are both variables, ${\cal M}(R) = {\cal M}(E)$
    \item If $x$ is a variable and $y$ is a constant, ${\cal M}(R) = {\cal M}(E)\langle y\rangle$
    \item If $x$ is a constant and $y$ is a variable, ${\cal M}(R) = \langle x\rangle{\cal M}(E)$
    \item If $x$ and $y$ are both constant, ${\cal M}(R) = \langle x\rangle{\cal M}(E)\langle y\rangle$
\end{enumerate}

\begin{figure*}[tb]
\centering
\includegraphics[width=0.8\textwidth]{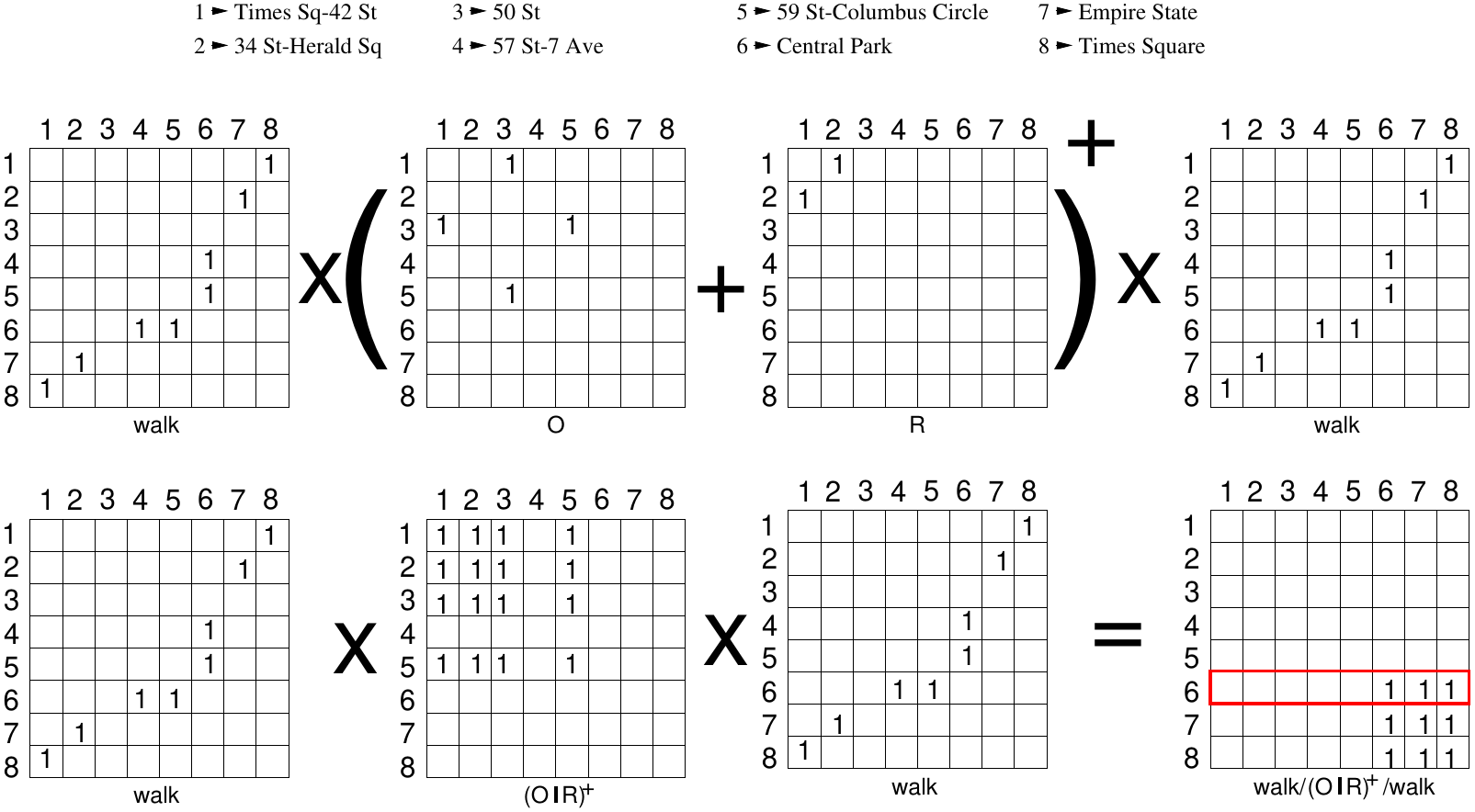}
\caption{Example of transforming the 2RPQ $R=(\textsf{Central Park}, \mathsf{walk}/(\mathsf{O} | \mathsf{R})^+/\mathsf{walk},\textsf{?y})$ into matrix operations. For readability we depict each node (station or point of interest) as an integer. On top, we can see the matrices and operations used to solve $E=\mathsf{walk}/(\mathsf{O} | \mathsf{R})^+/\mathsf{walk}$. Below, the second matrix represents ${\cal M}((\mathsf{O} | \mathsf{R})^+)$, which was obtained by computing the transitive closure on the sum of ${\cal M}(\mathsf{O})$ and ${\cal M}(\mathsf{R})$. The fourth matrix below is ${\cal M}(E)$. In order to solve $R$, as $x$ is the node $\textsf{Central Park}$, we restrict to the sixth row of ${\cal M}(E)$ and the solutions are $\textsf{Central Park}~(6)$, $\textsf{Empire State}~(7)$ and $\textsf{Times Square}~(8)$.}
\end{figure*}

\section{Implementation of the Boolean Matrix Algebra}

We now describe how the Boolean-matrix operations are carried out. To analyze the corresponding algorithms, we use $|M_{p}|$ as the number of 1s in the matrix, which is the number of edges with label $p$ in graph $G$. 
We represent each matrix $M_p$ using a $k^2$-tree of $\log_2 |V|$ levels, and each 1 in $M_p$ induces at most $\log_2 |V|$ 1s in its $k^2$-tree representation. As explained, per this representation we will assume $|V|$ is a power of 2. We will also use $v=|V|$, as well as $a=|A|$ and $b=|B|$ for the number of 1s in matrices $A$ and $B$.

We implement $k^2$-trees, and thus bitvectors with $\rank$ support, in C. We store the bitvector as consecutive bits packed in a 64-bit-words array. To support $\rank$ we store the cumulative sum of 1s up to every $s$th cell of the array. To save space, full 64-bit integers store the full sum only every $2^{16}$ bits, and the others are stored in relative form using 16-bit integers. To compute $\rank$ we start from the last recorded sum and use {\em popcount} on the full words until reaching the desired one, and a partial {\em popcount} on the desired word. Here $s$ allows trading space for time: we use $n/1024+n/(4s)$ additional bits of space for storing a bitvector $B[1\dd n]$, and compute $\rank$ in time $O(s)$. We use $s=4$.

In the sequel we describe how the different operations of the Boolean algebra are implemented on this representation. Transpositions are described immediately because they are incorporated to the data structure rather than executed as an operation; later we describe how the operations handle matrices marked as transposed. For every operation we also consider two aspects: (1) how to incorporate parallelism and (2) how to handle restrictions. 

Parallelism will be implemented with multithreading, but for simplicity it will be analyzed in the PRAM model of computation, assuming that the 1s are uniformly distributed on the matrices.\footnote{The actual multithreading adapts better to nonuniform distributions than our analysis under the PRAM model.} We will also assume that the number $p$ of processors is small compared to the number of the 1s and to the side of the matrices, which is realistic in multicore architectures. 

Restrictions indicate that we only want to retrieve a column or a row of the matrix after the operations, or even just a cell. A naive way to implement them is to first obtain the full matrix $M$ and then traverse the desired row or column. Yet, restrictions give an important opportunity of optimizing all the other operations. 

We show next how we extend $k^2$-trees to implement transpositions. Then, in Sections \ref{sec:sum} through \ref{sec:transitive-closure-implementation} we implement and analyze the main operations, namely sum (and relatives), multiplication, and transitive closure, respectively, on the extended format. 

Table~\ref{tab:times} shows the simplified time complexities we will obtain, and compares them with those of the baseline we describe in Section~\ref{sec:baseline} (which uses considerably more space). Note that the baseline time complexities are always smaller because $m \le v^2$.

\begin{table*}[t]
\caption{Simplified average time complexities for the main operations on $v\times v$ matrices with $m$ uniformly distributed 1s. The transitive closure assumes the result has $m^+$ 1s. The PRAM time uses $p \le m$ processors.}
\label{tab:times}
\centering
\begin{tabular}{l@{~}|@{~}c@{~}|@{~}c@{~}|@{~}c}
Operation & Baseline time & $k^2$-tree time & $k^2$-tree parallel time \\
\hline
Transposition     & $1$ & $1$ & $1$ \\
Sum and relatives & $m$ & $m \log v$ & $(1/p) m\log v + m + \log v \log p$ \\
Multiplication    & $(m^2/v) \log v$ & $m^{3/2}\log v$ & $(1/p)m^{3/2}\log v + m^2/v + \log v \log p$\\
Transitive closure & $(m^+)\log v$ & $(m^+)^{3/2}\log v$ & $(1/p)(m^+)^{3/2}\log v + (m^+)^2/v + \log v^2 \log p$ \\
\end{tabular}
\end{table*}

\subsection{Transposition}

Transposition is used to implement reversed edges, as seen in Section \ref{sec:evaluating-rpqs}. Instead of materializing the transposed matrix as a $k^2$-tree, we note that %$A^T = {A_0^T~A_2^T \choose A_1^T~A_3^T}$.
$$ A^T =
\Biggl(\mkern-5mu
\begin{tikzpicture}[baseline=-.65ex]
\matrix[
  matrix of math nodes,
  column sep=1ex,
] (m)
{
A_0^T & A_2^T \\
A_1^T & A_3^T\\
};
\draw[dotted]
  ([xshift=0.5ex]m-1-1.north east) -- ([xshift=0.5ex]m-2-1.south east);
\draw[dotted]
  (m-1-1.south west) -- (m-1-2.south east);
\end{tikzpicture}\mkern-5mu
\Biggr).$$
So, the $k^2$-tree for $A^T$ can be obtained by interchanging the roles of the second and third children of every node. We do not materialize this interchange,  but associate a {\em transposed} flag to every matrix, so we simply have to toggle it in order to transpose the matrix in $O(1)$ time.

\section{Boolean Sum and Relatives} \label{sec:sum}

In this section we address the set-like Boolean operations, with special emphasis on the Boolean sum (or disjunction) for its impact on later operations like multiplication. We start from known techniques \cite{QFPLG19}, and then improve the algorithms and adapt them to handle transpositions, parallelism, and restrictions. We also provide improved time complexity analyses.

If neither $A$ or $B$ is transposed, we can compute the sum $A+B$ with a simple sequential pass over both $k^2$-tree bitvectors \cite{QFPLG19}, merging their corresponding nodes levelwise without need of any $\rank$ operation. 
We implement this traversal with a queue of tasks, which are of two types. (\emph{1}) A {\em copy} task indicates to copy the next node from $A$ or $B$; and (\emph{2}) a {\em merge} task indicates merging the next nodes of $A$ and $B$. The queue is initialized with a merge task on both root nodes, the read-pointers (which indicate the next $k^2$-tree node to be read) at the beginning of the bitvectors of $A$ and $B$, and the write-pointer at the beginning of the output $k^2$-tree bitvector.

\begin{figure*}[t]
\centering

\includegraphics[width=0.8\textwidth]{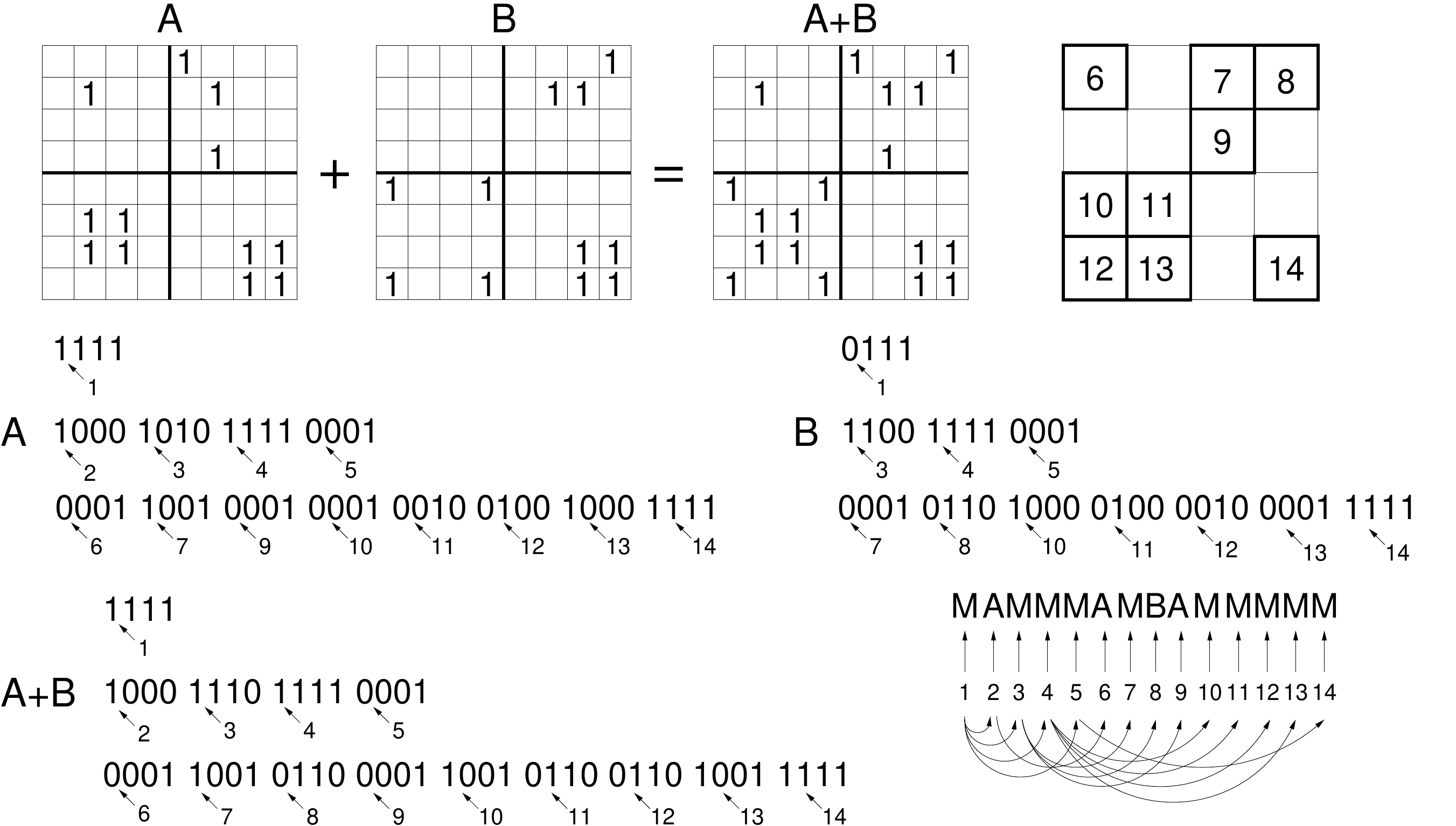}
\caption{Our sequential algorithm when summing matrices $A$ and $B$ shown on the top left. The $k^2$-tree representations of $A$, $B$, and $A+B$ are shown below the matrices, artificially separating the three levels in three lines. On the bottom left, the elements inserted in the queue along time, using \textsf{M} for ``merge'' and \textsf{A}/\textsf{B} for ``copy $A/B$''. The algorithm runs along 14 steps, one per element in the queue. Each such number in the queue has forward arcs towards the elements its step inserts; for example step 1, which corresponds to the whole matrix, inserts the elements 2 to 5 corresponding to the four quadrants. The scheme on the top right shows to which $2\times 2$ submatrices do the steps 6--14 correspond. The diagonal arrows in the $k^2$-tree representations of $A$ and $B$ show the position of the read-pointers at each step, and those on $A+B$ show the corresponding write-pointers. For example, in step 1, we read \textsf{1111} from $A$ and \textsf{0111} from $B$, thus we insert a ``copy $A$'' and three ``merge'' elements in the queue. Step 2 reads ``copy $A$'' from the queue and thus copies \textsf{1000} from the sequence of $A$ and appends it to that of $A+B$. Step 3 reads ``merge'' and thus merges \textsf{1010} from $A$, \textsf{1100} from $B$, 
 and writes \textsf{1110} to $A+B$. And so on.}
\label{fig:sum1}
\end{figure*}

To process a copy task, we append the next signature pointed by the read-pointer (of $A$ or $B$) to the output, and enqueue its (up to) 4 children as copy tasks for $A$ or $B$, respectively. To process a merge task, we append to the output the bitwise-or of the next 4-bit signatures pointed by the read-pointers of $A$ and $B$, and enqueue up to 4 new elements, as follows. For $i$ from 1 to 4, if the $i$th bit of the signatures of both $A$ and $B$ are 1, we append a merge task. If only one of them is 1, we append a copy task for the corresponding matrix. If none is 1, we do not append any task.
We do not append new tasks when the corresponding nodes are $k^2$-tree leaves. The process finishes when the queue becomes empty. Figure~\ref{fig:sum1} illustrates the algorithm. 

The total time is proportional to the sum of the number of nodes of both $k^2$-trees, $O(a\log (v^2/a)+b\log(v^2/b)) \subseteq O((a+b)\log v)$. We introduce a speedup that does not change the complexity but has a significant impact in practice: we do not append consecutive copy tasks for $A$ or for $B$ in the queue, but rather merge them into a single task that copies several signatures together, using a constant number of operations on computer words.

\subsection{Handling transpositions} 

If both $A$ and $B$ are transposed, we just merge them as described and mark the result as transposed. When one is transposed and the other is not, we cannot anymore resort to a sequential traversal of both bitvectors. Instead, we handle the sum as any other set-like operation, see next.
%The transposed one must already have $\rank$ support built to enable $k^2$-tree traversals. We traverse sequentially the non-transposed $k^2$-tree, and include in the queue the corresponding node of the transposed one (as those nodes are not read in left-to-right order). To generate the new tasks, we must use the $k^2$-tree traversal operations to locate the corresponding nodes in the transposed $k^2$-tree. 

%While the time complexity is the same as before, summing a transposed with a non-transposed matrix is slower in practice. We always choose that the transposed matrix is the one with a shorter bitvector (we can because $A^T+B = (A+B^T)^T$), in order to minimize the non-local traversals.

\subsection{Set-like operations} \label{sec:relatives}

\begin{figure*}[t]
\centering
\includegraphics[width=0.7\textwidth]{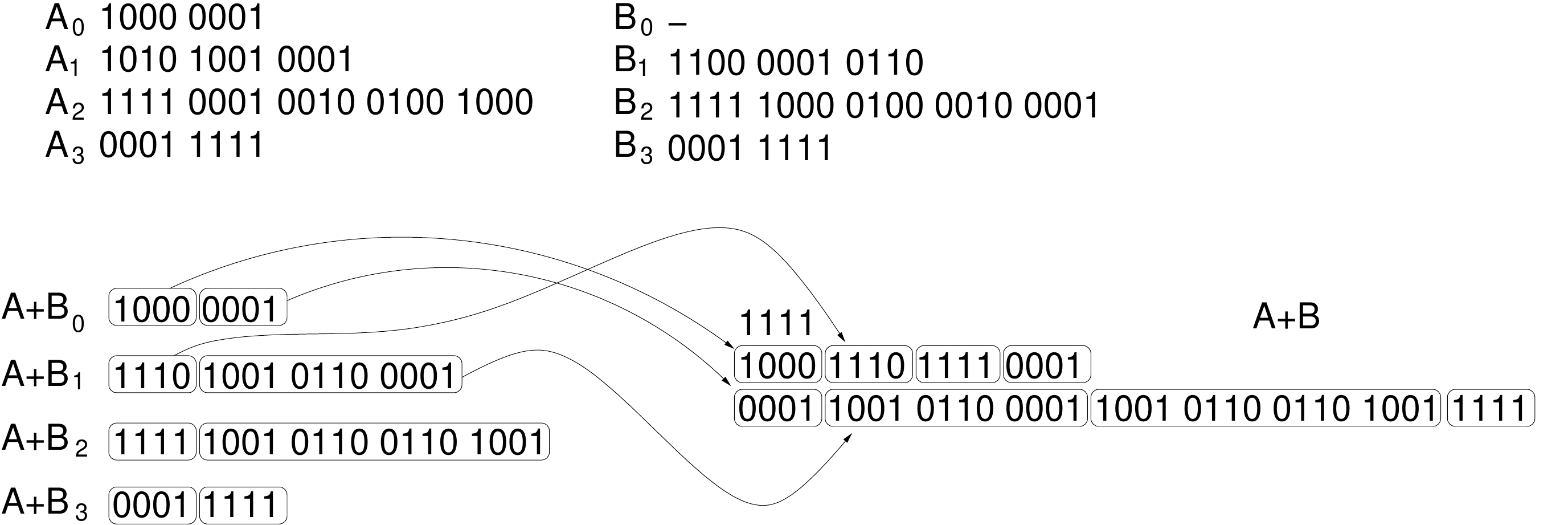}
\caption{The recursive algorithm to solve the same sum $A+B$ as in Figure~\ref{fig:sum1}. We set up the output space (bottom right of the figure) and then the $k^2$-tree bitvectors resulting from each quadrant are written directly to the corresponding levels as they are generated. The arrows show how this works for the quadrants $0$ and $1$. Because no quadrant is empty, the signature of the root of $A+B$ is \textsf{1111}.}
\label{fig:sum3}
\end{figure*}

Several other operations of the Boolean algebra have the same structure of the sum $A+B = A \cup B$ (i.e., Boolean ``or'' of the 1s): intersection $A \cap B$ (Boolean ``and'' of the 1s), difference $A - B$ (Boolean ``and not'' of the 1s), and symmetric difference $A\oplus B $ (Boolean ``exclusive or'' of the 1s). In general, those cannot be solved with the merge-like algorithm we described for the sum because they lack the key property that the signature of the resulting $k^2$-tree root is a function of the signatures of the $k^2$-tree roots of the operands (in the case of the sum, it is the bitwise-or of the signatures). Further, they may require skipping large submatrices of the operands. Instead, we must first operate the submatrices and only then define the signature of the result based on which are nonempty. We then resort to a recursive algorithm of the form

\begin{equation} \label{eq:matrix-op}
A \circ B = 
\Biggl(\mkern-5mu
\begin{tikzpicture}[baseline=-.65ex]
\matrix[
  matrix of math nodes,
  column sep=1ex,
] (m)
{
A_0 \circ B_0 & A_1 \circ B_1 \\
A_2 \circ B_2 & A_3 \circ B_3 \\
};
\draw[dotted]
  ([xshift=0.5ex]m-1-1.north east) -- ([xshift=0.5ex]m-2-1.south east);
\draw[dotted]
  (m-1-1.south west) -- (m-1-2.south east);
\end{tikzpicture}\mkern-5mu
\Biggr),
\end{equation}
where $A = { A_0~A_1 \choose A_2~A_3}$ and $B = { B_0~B_1 \choose B_2~B_3}$ are the submatrices into which the $k^2$-tree representation splits $A$ and $B$, and $\circ \in \{ \cup, \cap, -, \oplus \}$.
That is, we recursively operate the submatrices $A_i$ and $B_i$, for $0 \le i < 4$, obtaining up to $4$ submatrices $A_i \circ B_i$ represented as $k^2$-trees. Instead of producing the $k^2$-trees and later concatenate them levelwise, we prepare the memory space for the output separated by levels, so that the recursive calls directly append their results in the corresponding levels \cite{QFPLG19}. This works because a recursive tree traversal corresponds to a left-to-right traversal within each level. Figure~\ref{fig:sum3} illustrates the algorithm for the sum (or union), to allow contrasting with the previous algorithm in Figure~\ref{fig:sum1}.

An important improvement we make on top of the basic recursive algorithms \cite{QFPLG19} is that, when one of the two arguments is an empty submatrix, we may have to copy the other argument to the output. This occurs for both $A$ and $B$ in $A \cup B$ and $A \oplus B$, and for $A$ in $A-B$. Instead of carrying out this copy node by node of the $k^2$-trees, we perform a levelwise copy. In this copy we work $O(1)$ time per computed word copied, which in the transdichotomous RAM model of computation stores $\Theta(\log v)$ nodes of the $k^2$-tree. For example, copying a whole $k^2$-tree of $a$ leaves (and $O(a \log v)$ nodes) takes time $O(a+\log v)$, not $O(a \log v)$ (the second additive term stands for the $O(1)$-time overheads at each level). The impact of this improvement is made clear soon in the analysis and later in the experimental results.

\paragraph{Transpositions.}
When solving operations in this way, transpositions are handled easily by exchanging the meaning of $M_1$ and $M_2$ in every node of the $k^2$-tree bitvector, if $M = { M_0~M_1 \choose M_2~M_3}$ is transposed. As explained before, we use this technique for the sum when one matrix is transposed and the other is not. Otherwise, both the merge-like and the recursive algorithm can be used. The next analysis and later the experiments shed light on which algorithm is to be preferred depending on the case.

\paragraph{Analysis.}

Just as for the merge-like algorithm for $A+B$, the time complexity of the recursive algorithms is in $O(a\log (v^2/a)+b\log(v^2/b)) \subseteq O((a+b)\log v)$, for all the operations, as we work at most $O(1)$ time per node of the input and output $k^2$-trees. This analysis can be refined, however.

Let us start with the intersection, $C = A \cap B$. A first refinement is that there are at most $\min(a,b)$ elements carried to the output, not $a+b$, thus its time complexity is in $O(\min(a,b)\log v)$ because the algorithm traverses only the $k^2$-tree nodes below which both $A$ and $B$ have leaves. We can prove even more refined adaptive bounds by relating this problem to the adaptive intersection of integer sets \cite{BK08,DLM00}, in particular with the trie approach by Arroyuelo and Castillo \cite{ACcpm23}. Let $c=|C| \le \min(a,b)$ be the output size. We define conceptual integer sets $S_A$ and $S_B$, which represent the set of positions with 1s in the matrices $A$ and $B$, respectively. Therefore, the set $S_A \cap S_B$ represents the set of 1s in the matrix $A\cap B$. 
%The universe of these sets should be such that every possible entry in $A$ (resp.~$B$) has a universe element associated with it. As there are $v^2$ possible entries in each such matrix, both 
The sets $S_A$ and $S_B$ are subsets of the universe $[0{..} v^2)$, concretely 
%the matrix cell position $(i, j)$, for $0 \le i, j < v$, shall be represented by the universe element $\textsf{z-order}(i, j)$. That is, the order of the universe elements corresponds to the z-order of the matrix cells. Then, we define set 
$S_A = \{ \textsf{z-order}(i, j),~ A[i, j] = 1\}$, and analogously for $S_B$. As a consequence, the order of the elements in $S_A$ corresponds to a left-to-right traversal of the corresponding $k^2$-tree leaves, and therefore the algorithm we propose to compute $A\cap B$ mimics the set intersection algorithm proposed by Arroyuelo and Castillo \cite{ACcpm23}. The only difference is that this time the sets are represented as 4-ary tries (i.e., $k^2$-trees) rather than binary tries, but the properties needed to prove their adaptive bound still hold.  Then, the time for computing $A\cap B$ is $O(\delta \log(v^2/\delta))$, where $\delta$ is the {\em alternation measure} of $S_A$ and $S_B$ defined by Barbay and Kenyon \cite{BK08}, which in particular satisfies $c \le \delta \le \min(a,b)$. To this time, which measures the number of $k^2$-tree nodes traversed, we should add the $O(c\log (v^2/c)) \subseteq O(c\log v)$ cost to copy the nodes in the paths toward the $c$ resulting points, but this is subsumed by $O(\delta\log(v^2/\delta)) \subseteq O(\delta\log v)$.

Measure $\delta$ enables a finer analysis of the intersection time. It measures the number of times we {\em need} to switch between $S_A$ and $S_B$ in order to collect all the $S_A \cup S_B$ integers, along a left-to-right traversal of both ordered sequences. For example, if all the $a$ 1s of $A$ are in the first quadrant, and all the $b$ 1s of $B$ are in the third quadrant, then their corresponding measure is $\delta=O(1)$ because all the values in $S_A$ precede those in $S_B$, and thus we need $O(1)$ switches to collect them all. This yields an upper bound of $O(\log v)$ for our intersection. In this case the bound is pessimistic because our algorithm actually runs in $O(1)$ time, but it would be tight if the two clusters would split only in the same subgrid of depth $\Theta(\log v)$ of $A$ and $B$.

For the set difference, $C=A-B$, we note that $c \le a$, thus the time is in $O(a\log(v^2/a)) \subseteq O(a\log v)$. Further, since $A-B = A \cap \overline{B}$, we can reuse the analysis of the intersection to obtain a finer measure. Let $|\overline{B}| = \overline{b} = v^2 - b$ denote the number of elements in $\overline{B}$. The analysis proceeds as before, 
%defining sets $S_A$ and $S_{\overline{B}} = \{\textsf{z-order}(i, j)~|~0\le i, j < v~\wedge~B[i, j] = 0\}$, which has $|S_{\overline{B}}| = \overline{b}$ elements. Hence, computing $A-B$ is equivalent to computing the set intersection $S_A\cap S_{\overline{B}}$. Indeed, our algorithm behaves similarly as Arroyuelo and Castillo's algorithm \cite{ACcpm23}, in our case using a 4-ary $K^2-tree$. Thus, the running time of our approach is 
obtaining time $O(\overline{\delta}\log(v^2/\overline{\delta}))$, where $\overline{\delta}$ is the alternation measure of $S_A$ and  $S_{\overline{B}}$, which satisfies $c \le \overline{\delta} \le \min{\{a, \overline{b}\}}$.

Thanks to our improved algorithm to copy whole submatrices, we can also use the alternation measure to refine the time complexity $O((a+b)\log v)$ of the union and symmetric difference. Returning to our example where all the 1s of $A$ are in the first quadrant and all those of $B$ are in the third, our algorithms run in time $O(a+b+\log v)$. In general, all the $s$ integers of $S_A$ and $S_B$ that lie between two consecutive switches between $S_A$ and $S_B$ can be copied computer-word-wise, in time $O(s/\log v+1)$. As we have to traverse $O(\delta\log(v^2/\delta))$ nodes and to copy $O(c\log(v^2/c))$ nodes (where this time $\delta$ can be smaller than $c \le a+b$), we have a total time of $O(\delta\log(v^2/\delta)+c\log(v^2/c)/\log v) \subseteq O(a+b+\delta\log v)$ for the whole process.

We expect our refined analysis to show up in practice when the matrix densities are very different or they distribute non-uniformly.

\subsection{Parallelism} \label{sec:parsum}

Our basic merge-based summation algorithm described at the beginning of the section is difficult to parallelize, because we do not know where to start copying each of the summands at each level. The standard parallel merging algorithms, which first use a parallel-prefix scheme to find the cumulative sums of the arrays to merge and then copy each array independently, cannot be used here because our merge is not disjoint: there are some 1s at the same positions in $A$ and $B$. In order to determine where to copy the next subtrees, we must actually merge the current ones, and thus must proceed in sequence.

\begin{figure*}[t]
\centering
\includegraphics[width=0.8\textwidth]{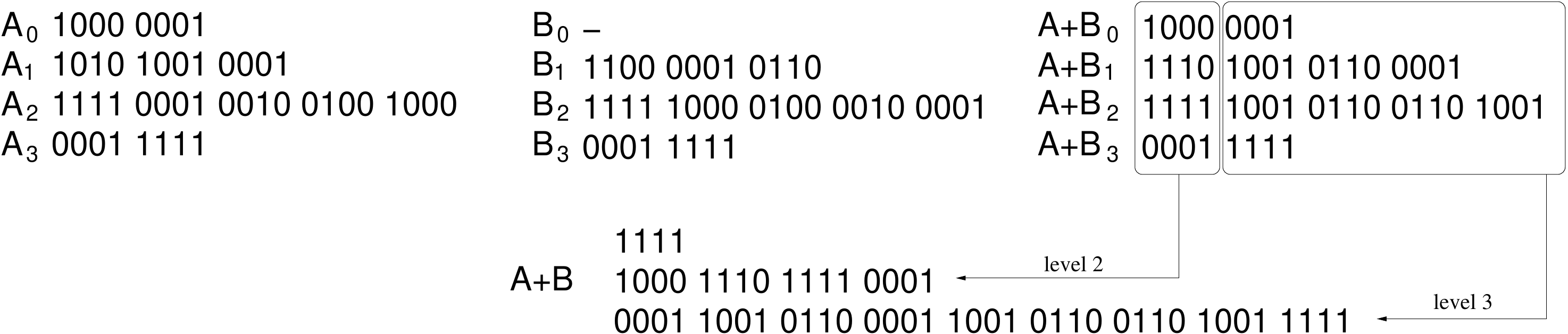}
\caption{The parallelizable recursive algorithm to solve the same sum $A+B$ as in Figure~\ref{fig:sum1}. We show the $k^2$-tree bitvectors of all the intervening quadrants as single sequences, and that of the result with one level per line. The recursive calls were called in parallel and have already produced the four quadrants of the result, $(A+B)_j$ for $0 \le j < 4$. Because the four are nonempty, the signature of the root of $A+B$ is \textsf{1111}. Now we copy the first levels of $(A+B)_j$ into the second level of the result, \textsf{1000 1110 1111 0001}. Finally, we copy the second levels of $(A+B)_j$ into the third level of the result.}
\label{fig:sum2}
\end{figure*}

The generic recursive algorithm for operation $\circ$ of Eq.~(\ref{eq:matrix-op}) is more amenable to parallelization. We can proceed in parallel for each $A_i \circ B_i$, this time writing each result as an independent $k^2$-tree, and then concatenate them sequentially at the end. Writing the results directly to the output is not possible in the parallel context because, again, we do not know in advance where to write. We speed up the concatenations by maintaining, for each $k^2$-tree bitvector, $O(\log v)$ counters of the sizes of the $k^2$-tree levels. Figure~\ref{fig:sum2} illustrates the algorithm we parallelize.

For all the operators, when one of the submatrices is zero, the result is either zero or the other submatrix, which helps speed up the computation. In particular, we share the submatrix in the second case instead of generating a new copy of it. We also do not need to build the $\rank$ data structures until the end of the whole operation, because the concatenation operates sequentially over the matrices. 

We implement a multithreaded version of this algorithm, where the recursive calls keep opening new threads for a few levels to avoid saturating the system with many more processes than processors.

\paragraph{Analysis.}
Let us first analyze the sequential cost of this algorithm.
In general, the concatenation time is proportional to the number of nodes of the operated $k^2$-trees, which as explained add up to $O((a+b)\log v)$. However, we do not concatenate the bitvectors bit by bit, but rather by whole computer words. Because of bit alignment issues, copying a computer word requires up to two read and two write operations, which in the transdichotomous RAM model contains $\Theta(\log v)$ bits. As a result, the total copying time is $O(a+b+\log v)$, the last term accounting again for a constant additive penalty per $k^2$-tree level copied. 

Consider now the whole operation time. In the worst case, there are $a+
b$ points across all submatrices in each level of the recurrence, thus the $a+b$ term above adds up to $O((a+b)\log v)$ across the $O(\log v)$ levels of the recursion. The terms $\log v$, instead, add up to $O((a+b)\log^2 v)$ in the worst case, in which each of the $a+b$ points are isolated and copied individually across all the levels. 

In order to avoid the additive penalties leading the cost, we will use a special concatenation method that is $O(a)$ time when we merge one submatrix of $a$ points with other three empty submatrices. Note that the resulting $k^2$-tree is identical to that of the nonempty submatrix preceded with a signature for the new root. We first write the new root signature, which marks the nonempty submatrix, and then concatenate the bitvector of that submatrix. The submatrix has $O(a\log v)$ bits but, as explained, we copy it by chunks of $\Theta(\log v)$ bits, so the total time is $O(a)$. Further, the $O(\log v)$ level counters of the nonempty submatrix are not copied, but shared and extended with one further entry. We call this a {\em trivial concatenation}, and say that it poses an additive penalty of $O(1)$ (as opposed to the $O(\log v)$ penalty posed by the nontrivial concatenations).

As said, we can have $O((a+b)\log v)$ concatenations along the process, but only $O(a+b)$ of those are nontrivial (those can be regarded as the branching nodes of a 4-ary tree with $a+b$ leaves). Nontrivial concatenations pose the $O(\log v)$ additive penalty, but that of trivial ones is only $O(1)$. Overall, the total time spent on additive penalties is $O((a+b)\log v)$, and this is also the total time complexity of the operations. 
 
We now give a PRAM-based analysis assuming that $p \le \min(a,b)$ and that the $a$ 1s in $A$ and the $b$ 1s in $B$ distribute uniformly. Let $S(v^2,a+b,p)$ be the parallel time on $v \times v$ matrices, $a+b$ points in total, and $p$ processors. Assuming we assign $p/4$ processors to build each submatrix $A_i \circ B_i$, the recurrence for the parallel time is 
\begin{eqnarray*}
&& S(v^2,a+b,p) ~=~ \\
&& S(v^2/4,(a+b)/4,p/4) + (a+b+\log_4(v^2)),
\end{eqnarray*}
where the second term (using convenient constants) stands for the cost to sequentially concatenate the $4$ resulting submatrices. The recursion continues up to the level $\ell' = \log_4 p$, where $4^{\ell'} \ge p$. At this level, the $a/p$ and $b/p$ points in each pair of submatrices are sequentially merged, in time $S(v^2/p,(a+b)/p,1) = O(\frac{1}{p}(a+b)\log v)$. The whole recurrence then solves as follows:\footnote{Note that all the concatenations up to level $\log_4 p$ are nontrivial, per our assumptions on the number of points in the submatrices.}
\begin{eqnarray} 
&& S(v^2,a+b,p) ~=~ \label{eq:parsum} \\ %S(v^2/4,(a+b)/4,p/4) + a+b+\log_4(v^2) \\
         %&=& 
&& S(v^2/p,(a+b)/p,1) + \sum_{\ell=0}^{\log_4 p-1} \left(\frac{a+b}{4^\ell} + \log_4\frac{v^2}{4^\ell}\right) \nonumber \\
&& ~=~ O\left(\frac{1}{p} (a+b)\log v + (a+b) +\log v \log p)\right). \nonumber
\end{eqnarray}
Although the analysis is simplified, it suggests we can expect a nearly perfect speedup, at least for small enough $p = O(\log v)$, and disregarding the small $O(\log v \log p)$ additive penalty.

\subsection{Restrictions} 

For $\langle r\rangle (A+B)\langle c\rangle $ (where it may be that only $\langle r\rangle $ or only $\langle c\rangle $ are present), we restrict the traversal of both matrices, acting as if the submatrices not intersecting the desired row and/or columm were empty. That is, we implement the restricted sum as $\langle r\rangle A\langle c\rangle  + \langle r\rangle B\langle c\rangle $. The merge-like algorithm for the sum cannot be used, however, because just as it happened for the generic operations $\circ$, we do not know beforehand whether a submatrix (or the merge of two) will be nonempty after restricting it to some row/column, even if it intersects the row/column. 

We then implement all the restricted operations, including the sum, using the recursive algorithms.
The only difference is that, when the row and/or column are restricted, only two of the four submatrices will be nonempty, and when both are restricted, only one submatrix will be nonempty.
%Letting $c$ be the number of points in the restricted matrix, if only a row or a column is restricted (so $c \le \min(v,a+b)$), we recurse on two submatrices, thus the recurrence for the time complexity is $T(v^2) = 2 \cdot T(v^2/4) + c + \log v$. Considering as before the worst case where we recurse for $\log_2 c$ levels until every submatrix has a single element, this solves to $T(v^2)=O((c+\log (v^2/c))\log v) \subseteq O((c+\log v)\log v$. This can be reduced to $O(c\log v)$ by merging the lower submatrices only every $\frac{1}{2}\log v$ levels. If both row and column are restricted, the time is just $O(\log v)$.

\section{Boolean Multiplication} \label{sec:mult}

For the multiplication $A \times B$ we use the following classic divide-and-conquer recursive procedure. Letting $A = { A_0~A_1 \choose A_2~A_3}$ and $B = { B_0~B_1 \choose B_2~B_3}$ as before, we recursively compute 8 products of those submatrices in order to produce
%$$ A \times B ~~=~~ { A_0 \times B_0 + A_1 \times B_2 ~~~ A_0 \times B_1 + A_1 \times B_3
%                      \choose
%                     A_2 \times B_0 + A_3 \times B_2 ~~~ A_2 \times B_1 + A_3 \times B_3 }.
%$$
\begin{eqnarray} 
&& A \times B = \label{eq:matrix-multiplication} \\
&& \Biggl(\mkern-5mu
\begin{tikzpicture}[baseline=-.65ex]
\matrix[
  matrix of math nodes,
  column sep=1ex,
] (m)
{
A_0\times B_0 + A_1\times B_2 & A_0\times B_1 + A_1\times B_3 \\
A_2\times B_0 + A_3\times B_2 & A_2\times B_1 + A_3\times B_3 \\
};
\draw[dotted]
  ([xshift=0.5ex]m-1-1.north east) -- ([xshift=0.5ex]m-2-1.south east);
\draw[dotted]
  (m-1-1.south west) -- (m-1-2.south east);
\end{tikzpicture}\mkern-5mu
\Biggr). \nonumber
\end{eqnarray}
A fortunate consequence of the $k^2$-tree representation is that, if any of those submatrices is empty (i.e., there is a 0 in the signature of the root of $A$ or $B$), then we know that its product with any other submatrix is also zero. Further, summing a product $A_i \times B_j$ with a zero matrix does not even need to copy the product; we just reference it as the final result, as explained.

Once the $k^2$-tree bitvectors of the four submatrices are recursively obtained, we concatenate them levelwise, as for our improved recursive sum operation. There is no need to build the $\rank$ data structures for this concatenation because it proceeds left-to-right in each level. We also maintain the $O(\log v)$ level counters in each $k^2$-tree to speed up concatenations.

Transpositions are again handled by exchanging the meaning of $M_1$ and $M_2$ in every node of the transposed matrices $M = { M_0~M_1 \choose M_2~M_3}$.

\subsection{Complexity} \label{sec:mult-anal}

One part of the multiplication cost is given by the number of recursive calls. We distinguish three stages to analyze a scenario with sparse matrices.
\begin{enumerate}
\item In the first stage, all the submatrices are nonempty. Since there are $4^\ell$ submatrices in level $\ell$, the worst case arises when every submatrix has points up to the level $\ell$ where we have $4^\ell \ge \min(a,b)$ submatrices, that is, up to level $\ell_1 = \log_4 \min(a,b)$. In this stage the cost follows the recurrence $T(v^2) = 8 \cdot T(v^2/4)$, therefore the cost up to level $\ell_1$ is $8^{\ell_1} = \min(a,b)^{3/2}$.
\item In the second stage, the worst case is that the emptier matrix has only one point in its submatrices while the fuller has $\max(a,b)/\min(a,b)$ evenly distributed points. This continues for $\ell_2 = \log_4 \frac{\max(a,b)}{\min(a,b)}$ further levels, in which the recurrence becomes $T'(v^2) = 2 \cdot T'(v^2/4)$ because the single point in the emptier submatrix can make us enter into at most two submatrices of the other. From each of the $8^{\ell_1}$ submatrices where stage 1 ends, we have then a cost of $2^{\ell_2} = \sqrt{\max(a,b)/\min(a,b)}$, which multiplied by $8^{\ell_1}$ yields the cost $\min(a,b)\sqrt{\max(a,b)}$ up to the end of stage 2.
\item In the third stage, we have just one point in each of the submatrices, so the cost is $\log_2 v - \ell_1-\ell_2 = \log_4(v^2/\max(a,b))$ to track a single point along both submatrices. This is done from each of the $8^{\ell_1} 2^{\ell_2} = \min(a,b)\sqrt{\max(a,b)}$ submatrices where stage 2 ends, leading to the final cost 
\begin{equation} \label{eq:multcalls}
O(\min(a,b)\sqrt{\max(a,b)}\log (v^2/\max(a,b))).
\end{equation}
\end{enumerate}

The second part of the multiplication cost is that of summing pairs of partial submatrices, recall Eq.~(\ref{eq:matrix-multiplication}). In the worst case, those matrices may add up to $a \cdot b$ points at across every level of the recursion. Just as in Section~\ref{sec:relatives}, where we had $a+b$ points in every level, the total merging cost of the partial results is $O(ab\log^2 v)$. With the technique of the trivial concatenations, this can be reduced to $O(ab \log v)$, where we pay $O(\log v)$ time only on the branching nodes of the resulting $k^2$-tree, which has $ab$ leaves at most. 

Note that this term dominates the cost of the first part. We can show that the average time, on matrices with uniformly distributed 1s, is better. We multiply $8^\ell$ pairs of $v/2^\ell \times v/2^\ell$ submatrices in level $\ell$. On average, each has $a/4^\ell$ 1s in $A$ and $b/4^\ell$ 1s in $B$. Every such $a_{ik}=1$ will pair with every such $b_{k'j}=1$ iff $k=k'$, which occurs with probability $1/(v/2^\ell)$, so on average there will be $8^\ell (a/4^\ell)(b/4^\ell)(2^\ell/v) = ab/v$. This leads to a total average time  of $O((ab/v)\log v)$ for the second part. Since $ab/v = \min(a,b)\max(a,b)/v \le \min(a,b)\sqrt{\max(a,b)}$ because $\max(a,b) \le v^2$, the bound $O(\min(a,b)\sqrt{\max(a,b)}\log v)$ of the first part of the cost dominates on the average.

\subsection{Parallelism}

A further advantage of Eq.~(\ref{eq:matrix-multiplication}) is that it is easily parallelized, as it features $8$ independent multiplications and $4$ sums, each sum depending only on the result of two multiplications. A multithreaded version assigns a thread to each of the 8 multiplications and to each of the 4 sums, forcing sequential execution of each sum after its two corresponding multiplications. The recursive calls are further parallelized for a few levels to avoid having many more processes than processors, as explained. 

We give a PRAM-based analysis of this process. Let $T(v^2,p)$ be the time of the algorithm on a $v \times v$ matrix and $p$ processors, where we assume that $p \le \min(a,b)$ and that the 1s distribute uniformly across the matrices. This implies that, with respect to the number of recursive calls, the parallelism is confined inside the stage 1 of the analysis in Section~\ref{sec:mult-anal}. 

We allocate $p/8$ processors to each of the $8$ multiplications. Then we will have more than one available processor per recursive call up to level $\ell_0 = \log_8 p$. From that level, each of the $8^{\ell_0} = p$ parallel calls start to run sequentially, for $\ell_1-\ell_0$ further levels. Each processor then runs $\ell_1-\ell_0$ levels of stage 1, then $\ell_2$ levels of the stage 2, and then the rest of the levels of stage 3. The total time spent by each processor is then $8^{\ell_1-\ell_0} 2^{\ell_2} \log(v^2/\max(a,b))$, which is exactly the
sequential cost of the recursive calls (Eq.~(\ref{eq:multcalls})) divided by $8^{\ell_0} = p$. In addition, we have the $O(\ell_0) = O(\log p)$ time spent in the first $\ell_0$ levels.

To anayze the 4 sums we can use Eq.~(\ref{eq:parsum}), replacing $a+b$ by $ab/v$. 
%For the first $\ell_0' = \log_4 p$ levels $0 \le \ell < \ell_0'$, assuming again that the (at most) $a \cdot b$ points are uniformly distributed, we perform all the sums of level $\ell$ in parallel time $S(v^2/4^\ell,(ab/v)/4^\ell,p/4^\ell)$. Adding up those levels we obtain
%$$\left(\frac{\log p}{p}+\frac{1}{\log v}\right)(ab/v)\log v+\log v \log^2 p.$$ 
%From level $\ell_0$ on, all the sums within each of the $p$ submatrices are sequential, but on average handle only $\min(v^2,ab/v)/p$ points. Therefore, each submatrix takes average time $O(\frac{1}{p}\min(v^2,ab/v)\log^2 v)$.
The total parallel time is then
\begin{eqnarray} 
&& T(v^2,p) ~=~ \label{eq:parmult} \\
&& \!\!\!\!O\left( %\log^2 p\log v + 
\frac{1}{p}\min(a,b)\sqrt{\max(a,b)}\log v + \frac{1}{p}\,\frac{ab}{v}\log v \right. \nonumber \\
&& \left. ~~~~~~~~ + \frac{ab}{v} + \log v \log p\right) ~=~ \nonumber \\
&& \!\!\!\!O\left( \frac{1}{p}\min(a,b)\sqrt{\max(a,b)}\log v + \frac{ab}{v}  + \log v \log p\right) \nonumber
\end{eqnarray}
which, compared to the sequential time, again suggests we can expect a nearly perfect speedup in our multithreaded implementation. If $p=O((v\log v)/\sqrt{\max(a,b)})$, in particular, the speedup is perfect except for the small additive term $O(\log v \log p)$.

\subsection{Restrictions}

A restricted product $\langle r\rangle  (A \times B) \langle c\rangle $ is handled as $(\langle r\rangle A) \times (B\langle c\rangle )$, where again only one of the restrictions may be present. We consider the column or row restrictions along the whole recursion, pretending that the submatrices that do not intersect the desired row or column are empty. 

Having one restriction (row or column) ensures that at most $6$ or the $8$ multiplications in Eq.~(\ref{eq:matrix-multiplication}) are nonzero, thereby modifying the recurrence of the number of multiplications to $T(v^2) = 6 \cdot T(v^2/4)$, which solves to $T(v^2) = O((\min(a,b)^{\log_4 6})$. Multiplied by the $2^{\ell_2}\log_4(v^2/\max(a,b))$ cost of stages 2 and 3 (where the restrictions yield no better upper bounds) we obtain the final bound,
$O(\min(a,b)^{\log_4 3}\sqrt{\max(a,b)}\log v)$; $\log_4 3 < 0.8$. 

For the sums, we note that on average only one out of $v$ of the $a$ 1s in $A$ or the $b$ 1s in $B$ satisfy the row or column restriction, so the average number of points to sum per level is $ab/v^2$. The time of the second part then becomes $O((ab/v^2)\log v)$. 

Having both row and column restrictions yields $T(v^2) = 4 \cdot T(v^2/4)$, which solves to $T(v^2) = O(\min(a,b))$ and to $O(\sqrt{ab}\log v)$ for the three stages, plus just $O(\log^2 v)$ time for the sums.

\section{Transitive Closure} \label{sec:transitive-closure-implementation}

A simple positive transitive closure algorithm obtains $A^+$ by iteratively computing 
$A \gets A + A \times A$ until no change occurs in $A$ \cite{Furman70}. This occurs at most after $\log_2 v$ iterations, so the time complexity is $O(\log v)$ times that of multiplying $A$ by itself. The non-positive transitive closure is computed as $A^* = I + A^+$, where $I$ is the identity matrix. Transposed matrices can be operated as is and the result would be transposed as well.

Since the number $a$ of 1s in $A$ grows in every iteration until reaching $a^+=|A^+|$, we can use Eq.~(\ref{eq:multcalls}) with $a=b=a^+$ to obtain an average time of $O((a^+)^{3/2}\log^2 v)$ along the $O(\log v)$ matrix multiplications, assuming a uniform distribution of the 1s. The non-positive closure adds $O((a^+ + v)\log v)$ further time for the final sum. 

We now introduce a more efficient algorithm, which obtains $A^+$ at the cost of $O(1)$ multiplications. Inspired by Warshall's algorithm \cite{War62} (and, in a way, in the ZCQ decomposition \cite{Penn06}), we compute $A^+$ in two steps. Let $A={A_0~A_1 \choose A_2~A_3}$ be a $v \times v$ matrix. In the first step we obtain the matrix $A'$, where $a'_{i,j}=1$ iff we can go from node $i$ to node $j$ through a single edge or using only intermediate nodes in $[1\dd v/2]$. This is computed as
\begin{eqnarray} 
&& A' ~=~ 
\Biggl(\mkern-5mu
\begin{tikzpicture}[baseline=-.65ex]
\matrix[
  matrix of math nodes,
  column sep=1ex,
] (m)
{
A_0' & A_1' \\
A_2' & A_3' \\
};
\draw[dotted]
  ([xshift=0.5ex]m-1-1.north east) -- ([xshift=0.5ex]m-2-1.south east);
\draw[dotted]
  (m-1-1.south west) -- (m-1-2.south east);
\end{tikzpicture}\mkern-5mu
\Biggr) ~=~
\label{eq:closure1} \\
&& 
\Biggl(\mkern-5mu
\begin{tikzpicture}[baseline=-.65ex]
\matrix[
  matrix of math nodes,
  column sep=1ex,
] (m)
{
A_0^+  & A_1 + A_0' \times A_1 \\
A_2 + A_2 \times A_0' & A_3 + A_2 \times A_1' \\
};
\draw[dotted]
  ([xshift=-0.5ex]m-1-2.north west) -- ([xshift=-0.5ex]m-2-2.south west);
\draw[dotted]
  (m-2-1.north west) -- (m-2-2.north east);
\end{tikzpicture}\mkern-5mu
\Biggr). \nonumber
\end{eqnarray}
(where $A_3'$ can also be computed as $A_3 + A_2' \times A_1$). So we first compute $A_0' = A_0^+$ recursively, then $A_1'$ and $A_2'$ (which depend on $A_0'$), and finally $A_3'$ (which depends on $A_1'$, or on $A_2'$ in its alternative formulation).

In the second step, we also permit the paths go through nodes in $[v/2+1\dd v]$, thereby completing the closure. The resulting matrix is computed as
\begin{eqnarray} 
&& A^+ = A'' = 
\Biggl(\mkern-5mu
\begin{tikzpicture}[baseline=-.65ex]
\matrix[
  matrix of math nodes,
  column sep=1ex,
] (m)
{
A_0'' & A_1'' \\
A_2'' & A_3'' \\
};
\draw[dotted]
  ([xshift=0.5ex]m-1-1.north east) -- ([xshift=0.5ex]m-2-1.south east);
\draw[dotted]
  (m-1-1.south west) -- (m-1-2.south east);
\end{tikzpicture}\mkern-5mu
\Biggr)
~=~ \label{eq:closure2} \\
&& \Biggl(\mkern-5mu
\begin{tikzpicture}[baseline=-.65ex]
\matrix[
  matrix of math nodes,
  column sep=1ex,
] (m)
{
A_0' + A_1' \times A_2'' & A_1' + A_1' \times A_3'' \\
A_2' + A_3'' \times A_2' & (A_3')^+ \\
};
\draw[dotted]
  ([xshift=0.5ex]m-1-1.north east) -- ([xshift=0.5ex]m-2-1.south east);
\draw[dotted]
  (m-1-1.south west) -- (m-1-2.south east);
\end{tikzpicture}\mkern-5mu
\Biggr). \nonumber
\end{eqnarray}
(where $A_0''$ can also be computed as $A_0' + A_1'' \times A_2'$). This time we start by computing $A_3'' = (A_3')^+$ recursively, then compute $A_2''$ and $A_1''$ (which depend on $A_3''$), and finally $A_0''$ (which depends on $A_2''$ or on $A_1''$).

\subsection{Correctness}

We call $X = [1\dd v/2]$ and $Y=[v/2+1\dd v]$ and use the notation $[M]_{i,j} = m_{i,j}$ for any matrix $M$. In $A'$, we must show that $[A']_{i,j}=1$ iff there exists a (nonempty) path from node $i$ to node $j$ whose sequence of intermediate nodes is in $X^*$ (the Kleene closure of $X$). It is easily seen that $A_0' = A_0^+$, because the allowed intermediate nodes are precisely all those in $A_0$. We inductively assume that $A_0' = A_0^+$ is computed correctly.

The cells $[A']_{i,j}$ that fall in $A_1'$ satisfy that $i \in X$ and $j \in Y$. Every path from $i$ to $j$ whose intermediate nodes are in $X$ corresponds to a path of zero or more edges starting at $i$ and ending in some $k \in X$, plus a final edge from $k$ to $j$. Therefore, either $[A]_{i,j}=1$ (i.e., $k = i$), or $[A']_{i,k}=1$ and $[A]_{k,j}=1$. This is equivalent to $[A_1 + A_0' \times A_1]_{i,j-n/2}=1$. The case of $A_2'$ is analogous: any path from $i \in Y$ to $j \in X$ through a sequence of nodes in $X^*$ starts with an edge from $i$ to some $k \in X$ and follows with a path (of length zero or more) from $k$ to $j$, thus it corresponds to $[A_2 + A_2 \times A_0']_{i-n/2,j}=1$.

Finally, a path from $i \in Y$ to $j \in Y$ that can have a sequence of intermediate nodes in $X^*$ can be either a direct edge from $i$ to $j$ with no intermediate nodes (for which we must have $[A_3]_{i-n/2,j-n/2}=1$), or it can be formed by an edge from $i$ to some $k \in X$ followed by a path from $k \in X$ to $j \in Y$ using intermediate nodes in $X$ (for which we must have $[A_2 \times A_1']_{i-n/2,j-n/2} = 1$). Thus we obtain $A_3' = A_3 + A_2 \times A_1'$. We can analogously derive the equivalent formula $A_3' = A_3 + A_2' \times A_1$.

For the second step, we start by computing $A_3'' = (A_3')^+$ recursively, and inductively assume its computation is correct. This corresponds to a concatenation of paths that start and end in $Y$, going through zero or more nodes of $X$ between each pair of nodes in $Y$, and where the final node in $Y$ of each path is the initial node of the next. The intermediate nodes then form a sequence $x_1 y_1 x_2 y_2 \ldots y_m x_{m+1}$, where $y_r \in Y$ and $x_r \in X^*$. The set of all those sequences is then $(X^*Y)^* X^*$. Therefore, $[(A_3')^+]_{i-n/2,j-n/2}=1$ whenever there is a path between $i$ and $j$ whose intermediate sequence of nodes is in $(X^*Y)^* X^*$. But then, note that $(X^*Y)^* X^* = (X|Y)^*$ is an easy to prove equality between regular languages. Thus, the path between $i$ and $j$ can use any number of intermediate nodes in $[1\dd n]$, and then $(A_3')^+ = A_3'' = (A^+)_3$.

Let us now consider the computation of $A_2''$. It should hold $[A_2'']_{i-n/2,j}=1$ iff there is a path from $i \in Y$ to $j \in X$ going through a sequence of zero or more intermediate nodes in $X$ or $Y$. If there are no nodes of $Y$ in such sequence, then it belongs to $X^*$ and it must hold $[A_2']_{i-n/2,j}=1$. Otherwise, let $y$ be the last node belonging to $Y$ in the sequence. The sequence is then of the form $z y x$, where $z \in (X|Y)^*$ and $x \in X^*$. Then it must hold that $[A_3'']_{i-n/2,y-n/2}=1$ and $[A_2']_{y-n/2,j}=1$, and consequently $[A_3'' \times A_2']_{i-n/2,j-n/2}=1$. The formula $A_2'' = A_2' + A_3'' \times A_2'$ is then proved. The case $A_1'' = A_1' + A_1' \times A_3''$ is analogous.

The final case, $A_0''$, is also analogous. It must hold that $[A_0'']_{i,j}=1$ iff there is path from $i$ to $j$ with intermediate nodes in $X$ or $Y$. If this sequence has only nodes in $X$, then it belongs to $X^*$ and it must hold that $[A_0']_{i,j}=1$. Otherwise, we can partition the sequence as $xyz$, where $x \in X^*$ and $y$ is now the first occurrence of an element in $Y$. Then it must be that $[A_1']_{i,y-n/2}=1$ and $[A_2'']_{y-n/2,j}=1$, and as a consequence it must hold that $[A_1' \times A_3'']_{i,j}=1$.

\subsection{Complexity}

Let $C(v^2)$ be the number of recursive calls to compute the closure of a $v \times v$ matrix, and $T(v)$ the number of recursive calls to multiply two $v \times v$ matrices. Our computation in Eqs.~(\ref{eq:closure1}) and (\ref{eq:closure2}) follows the recurrence
\begin{equation*}
C(v^2) = 2 \cdot C(v^2/4) + 6\cdot T(v^2/4).
\end{equation*}
%Indeed, we also perform 6 sums, which we disregard as their complexity is dominated by the multiplication cost. 
Note that, if we replace $C(v^2)$ by $T(v^2)$, we obtain the same recurrence of Section~\ref{sec:mult-anal}. Therefore, we can prove by induction on $v$ that $C(v^2) = T(v^2)$, and thus the number of calls in our closure algorithms is the same as in a multiplication. The number of elements in the matrices we multiply, however, can be as high as $a^+$. Using Eq.~(\ref{eq:multcalls}) we obtain $O((a^+)^{3/2}\log(v^2/a^+))$ recursive calls.

For the cost of the sums (both the ones done inside the multiplications and those of Eqs.~(\ref{eq:closure1}) and (\ref{eq:closure2})), we again assume that there are $(a^+)^2/v$ elements on average in every level. Since $a^+ \le v^2$, the total average cost of our transitive closure algorithm is 
\begin{equation} \label{eq:closure}
O\left((a^+)^{3/2}\log v\right).
\end{equation}
This is $\log v$ times less than the cost of the standard technique, and corresponds to multiplying two uniformly distributed matrices with $a^+$ 1s.

\subsection{Parallelism}

It can be seen in Eqs.~(\ref{eq:closure1}) and (\ref{eq:closure2}) that the computation of $A_2'$ can be carried out in parallel with those of $A_1'$ and $A_3'$, though $A_3'$ must be computed after $A_1'$ (and all must be computed after $A_0'$). Analogously, $A_1''$ can be computed in parallel with $A_2''$ and $A_0''$, all after $A_3''$. In addition, we can use parallelism to perform each isolated multiplication.

To analyze this process, let us call $T(v^2,a^+,p)$ the time for multiplying two random matrices with $a^+$ 1s on a $v \times v$ submatrix with $p$ processors. Combining Eqs.~(\ref{eq:parmult}) and (\ref{eq:closure}) and ignoring big-$O$ notation, we get 
$$ T(v^2,a^+,p) = \frac{1}{p}(a^+)^{3/2}\log v + 
                    (a^+)^2/v + \log v \log p.$$
Similarly, let $C(v^2,a^+,p)$ be the time to compute the closure on a $v \times v$ matrix $A$, assuming that there are already $a^+ = |A^+|$ uniformly distributed 1s in $A$, and with $p$ processors. Since we perform, on $v/2 \times v/2$ submatrices having on average $a^+/4$ elements, a sequence of two recursive calls and two multiplications, plus other two pairs of multiplications in parallel (i.e., those of $A_1'$ with $A_2'$ and those of $A_1''$ with $A_2''$), the recurrence for $C$ is
\begin{eqnarray*}
&& C(v^2,a^+,p) ~=~ \\
&& 2\cdot C(v^2/4,a^+/4,p) + 2\cdot T(v^2/4,a^+/4,p) + \\
&& 2\cdot T(v^2/4,a^+/4,p/2) ~\le~ \\
&& 2\cdot C(v^2/4,a^+/4,p) + \frac{3}{4}\cdot \frac{1}{p}(a^+)^{3/2}\log v + \\ && \frac{1}{2}(a^+)^2/v + 4 \log v \log p ~\le~ \\
&& \frac{1}{p}(a^+)^{3/2}\log v + \frac{2}{3} \cdot (a^+)^2/v + O(\log^2 v \log p),
\end{eqnarray*}
where the first two terms are obtained by unrolling the recurrence into exponentially decreasing terms, and the latter one is obtained by noting that 
the recursion ends at the level $\ell$ where $a^+/4^\ell=1$, and bounding $\log a^+ = O(\log v)$. The result is $O(T(v^2,a^+,p)+\log^2 v \log p)$, with the same constant in the leading term. This suggests we can expect a parallel time proportional to that of multiplying the resulting matrix, though this time the additive penalty may become noticeable.

\subsection{Restrictions}

Operation $A^+\langle c\rangle $ is implemented as
$S \gets (E + A)\langle c\rangle $, where $E$ is the empty matrix, and then repeatedly doing $P \gets A \times S$ and $S \gets S + P$ until $S$ does not change. Note that the only nonzero column of $P$ and $S$ is $c$. To implement $A^*\langle c\rangle $ we start with $S = (I + A)\langle c\rangle$ instead. 
A row restriction $\langle r\rangle A^+$ is handled analogously, starting with $S = \langle r\rangle (A + E)$ and then iterating over $P \gets S \times A$ and $S \gets S + P$, or using the initial step $S \gets \langle r\rangle  (I + A)$ for $\langle r\rangle A^*$. 

Note that, unlike the standard algorithm \cite{Furman70}, this iteration does not make the path lengths grow exponentially for the transitive closure, but linearly. Therefore, we could need up to $v$ iterations to compute the closure. In practice, the closure is reached much sooner and the operations are significantly faster, leading to a solution that is much faster than our new transitive closure algorithm. 

When both row and column are restricted, we only want a cell of the transitive closure. We then choose the row/column with fewer elements in $A$ and run a row-restricted or column-restricted closure, whichever is emptier. At each step, we check if the desired cell is full, stopping immediately if so.

\section{Query Plan}

\begin{figure*}[t]
\centering
\includegraphics[width=0.9\textwidth]{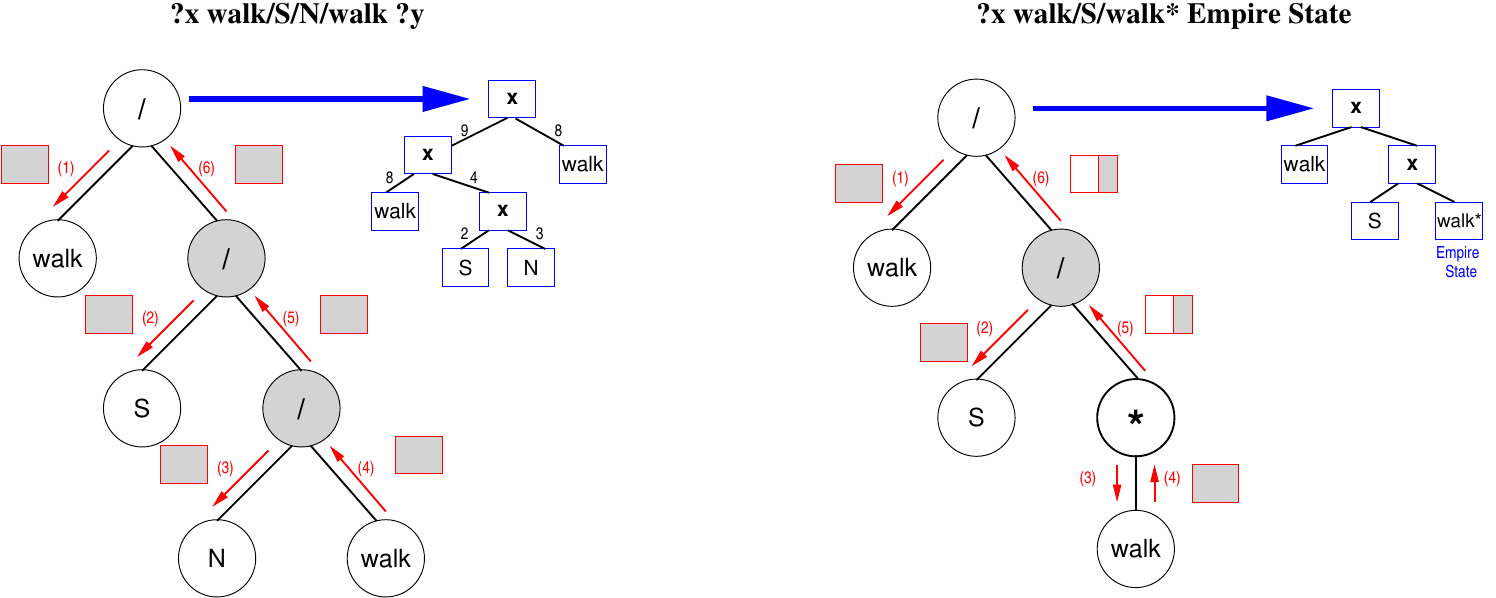}
\caption{Two examples for our query plan algorithm. For each query, the left part represents the syntax tree of the regular expression. The gray nodes depict delayed operations. The full- and partial-filled rectangles mean non-restricted and restricted operations, respectively. The tree pointed by the root represents the order of operations that the root has to compute, where the edges denote the size of the matrix representation. In $\textsf{?x}\ \mathsf{walk}/\mathsf{S}/\mathsf{N}/\mathsf{walk}\ \textsf{?y}$, the query is transformed into a sequence of multiplications on the matrices of the predicates without restrictions. Therefore, those multiplications are delayed until the root. In the root, we firstly choose the multiplication ${\cal M}(\mathsf{S}) \times {\cal M}(\mathsf{N})$ because the sum of their sizes is $5$, smaller than the sum of the first ($10$) and last pair ($11$). Then ${\cal M}(\mathsf{S})$ and ${\cal M}(\mathsf{N})$ are replaced by their product, of size 4, and the procedure continues recursively. Regarding  $\textsf{?x}\ \mathsf{walk}/\mathsf{S}/\mathsf{walk^*}\ \textsf{Empire State}$, we can restrict the column of our root. The complete right branch of the tree inherits that restriction, except the node $\mathsf{walk}$ due to the Kleene star operator. Once ${\cal M}(\mathsf{walk^*})$ is computed, we restrict it to the corresponding column. Since the previous operations are multiplications, ${\cal M}(\mathsf{S}) \times {\cal M}(\mathsf{walk^*})\langle  \textsf{Empire State} \rangle$ is delayed until the root. In contrast to the previous example, since the last matrix is restricted by a column, the algorithm computes the multiplication right to left.}
\label{fig:query-plan}
\end{figure*}

We first build the syntax tree of the 2RE $E$ of the 2RPQ $(x,E,y)$. In principle, we can simply traverse the syntax tree and solve it in postorder in the standard way, interpreting each leaf $p$ as the matrix $M_p$, $\rev p$ as $M_p^T$, and $\varepsilon$ as $I$, and interpreting the internal nodes as the corresponding operations on the matrices resulting from their children, according to the translations of Section~\ref{sec:evaluating-rpqs}. Our particular application, however, enables some relevant optimizations.

Let us first assume that both $x$ and $y$ are variables. A first simple optimization is that the closures are idempotent, so a sequence of closures is reduced to one. More precisely, $(A^*)^* = (A^*)^+ = (A^+)^* = A^*$ and $(A^+)^+ = A^+$. Sums and products yield more important optimizations, though. 
%Every regular expression $E$ can be represented using its syntax tree, a hierarchical binary tree structure that represents the relationships between its elements and operators. In the context of RPQs, each node $v$ stores an operator or a label, and its descending syntax (sub)tree describe the (sub)expression $E(v)$. Note that the root represents the whole expression $E$.
%
%\adrian{Figura con syntax tree}
%
%
% In order to solve $R=(x,E,y)$ our algorithm traverses the syntax tree of $E$ in postorder and translates each subexpression in Boolean-matrix operations according to the visited nodes and the translations of Section~\ref{sec:evaluating-rpqs}. 
%However, computing a subexpression $E(v)$ as soon as we solve its dependencies (i.e., every subexpression of the children of $v$ was solved) may not be a good option in certain cases because of its bad performance. For this reason, we propose a strategy that delays some operations until they are actually needed.
%
%Firstly, let us assume $R=(x, E, y)$ is not restricted, that is, $x$ and $y$ are variables. During the postorder traversal, the closures are solved once we reach an operator node because we cannot group them together. Note that having two consecutive closures in the regular expression, can be simplified to using the most restrictive one. With respect to the leaves, we just need to obtain the matrix of its label. Therefore, there are only two operations that can be delayed: sums and products. 

 \paragraph{Sums.} We exploit the fact that the Boolean sum is commutative and associative to carry out a sequence of consecutive sums, $E_1~| \dots|~ E_m$, in the best possible order. Since the cost of computing $A+B$ is proportional to $|A|+|B|$, if it were the case that $|A+B|=|A|+|B|$, the best possible order would be given by building the Huffman tree \cite{Huf52} of the matrices $A_i = {\cal M}(E_i)$ using $|A_i|$ as their weight. Since, instead, it holds that $\max(|A|,|B|) \le |A+B| \le |A|+|B|$, we opt for a heuristic that simulates Huffman's algorithm on the actual size of the matrices as they are produced. Concretely, we start with $\{ A_1, \ldots, A_m \}$ and iteratively remove from the set the two matrices $A_i$ and $A_j$ with the smallest sizes, sum them, and return $A_i + A_j$ to the set, until it has a single matrix.

%\paragraph{Sums.} The syntax tree can contain a node $v$ whose operator is $|$ and its $E(v)$ has the form $E_1~| \dots|~ E_m$, where $E_i$ is another regular expression. Note that ${\cal M}(E(v)) = {\cal M}(E_1) + \dots + {\cal M}(E_m)$. By following the postorder traversal we should solve ${\cal M}(E_{m-1})+{\cal M}(E_m)$ before the others, which can be expensive if the number of merge operations is large. Since the sum of matrices is commutative, we can delay that computation until we come back to $v$ and choose another order that minimizes the number of merge tasks. 
%
%The sum of two sparse matrices tends to be faster than in dense matrices because they need fewer merge tasks (e.g., the addition of an empty node with a non-empty one, requires a copy task). Therefore, we look for the two most sparse matrices $A$ and $B$ and compute $C=A+B$. Then, we replace both submatrices with $C$ and we repeat those steps until obtaining the final matrix, which is the result of $E(v)$. Detecting whose are the two most sparse matrices is easily computed by choosing those $A$ and $B$ that get the smallest $a$ and $b$, respectively.

\paragraph{Products.} Matrix multiplication is not commutative but still associative, so we can decide the order in which the sequence of multiplications to compute $E_1 ~/ \cdots /~ E_m$ is carried out. We cannot apply the well-known optimal algorithm to choose the order for dense matrices \cite[Sec.~15.2]{CLRS09} because the time complexity of our sparse matrix multiplications depends on the number of 1s in the matrices. Further, this number of 1s can increase or decrease after a multiplication. We then opt for a heuristic analogous to the one we use for sums: we start from the sequence $A_1,\ldots,A_m = {\cal M}(E_1),\ldots,{\cal M}(E_m)$ and iteratively choose the consecutive pair $A_i$, $A_{i+1}$ that minimizes $|A_i|+|A_{i+1}|$, multiply them, and replace the pair by $A_i \times A_{i+1}$, until the sequence has a single element.  

%\paragraph{Products.} Similar to the sums, we can find a node $v$ whose expression translation makes ${\cal M}(E(v)) = {\cal M}(E_1) \times \dots \times {\cal M}(E_m)$. Note that the product of two sparse matrices can generate a more sparse matrix. Therefore, delaying the products until we return to $v$ and multiplying first those that are the most sparse is a good option to solve $E(v)$ in an efficient way. 
%
%However, the product of matrices is not commutative, thus we need to respect the order of the matrices. For this reason, we look for the pair of consecutive matrices $A$ and $B$ that minimizes $\min(a, b)$. We compute the result $C=A \times B$, and replace both matrices with $C$. This procedure is recursively repeated until obtaining the final matrix.

\paragraph{Handling restrictions.}

When $x$ (resp., $y$) is a constant we are restricting a row (resp., column) of the matrix after the operations. For efficiency, then, we apply the restricted operations as described. Regarding the sums, because $\langle r\rangle (A+B) \langle c\rangle = \langle r\rangle A \langle c\rangle + \langle r\rangle B \langle c\rangle$, we can restrict all the involved matrices at the same time. Consequently, the sum can be computed in any order, and the plan still focuses on looking for the best order as described above. In the restriction on products, we obtain a sequence $\langle r\rangle A_1 \times \cdots \times A_m \langle c\rangle$ (where only $\langle r \rangle$ or only $\langle c \rangle$ could be present). Consider the case $\langle r\rangle A_1 \times \cdots \times A_m$. The number of 1s reduces faster when multiplying the pair that contains the restricted matrix, so we compute $A' = \langle r\rangle A_1 \times A_2$. The matrix $A'$ already has all zeros except in row $r$, so we can continue left-to-right in the sequence with normal matrix multiplications, $A' \times A_3$, and so on. The case $A_1 \times \cdots \times A_m \langle c\rangle$ is analogous, starting with $A' = A_{m-1} \times A_m \langle c\rangle$ and then completing the multiplications right to left. When both restrictions are present, we choose an end and proceed as explained until the final multiplication, $\langle r\rangle A' \times A'' \langle c\rangle$, which is carried out with the restricted multiplication algorithm to enforce the other restriction. 

Some restrictions can be inherited by the operands of a node, which speeds up processing. Since $\langle r\rangle (A+B) \langle c\rangle = \langle r\rangle A \langle c\rangle + \langle r\rangle B \langle c\rangle$, both children of a sum inherit the same restrictions. Instead, the product satisfies $\langle r\rangle (A \times B) \langle c \rangle = (\langle r\rangle A) \times (B \langle c \rangle)$, thus only the left child inherits a row restriction and only the right child inherits a column restriction. Closures do not inherit their restrictions to their operand, because  $\langle r \rangle A^* \langle c \rangle \neq (\langle r \rangle A \langle c \rangle)^*$ and $\langle r \rangle A^+ \langle c \rangle \neq (\langle r \rangle A \langle c \rangle)^+$. Restrictions are not inherited to leaves of the syntax tree, however, because internal operands handle them more efficiently than leaves. On the other hand, they are removed from parents when inherited to children because the nonrestricted operands run faster when their operands have already been restricted. 

Finally, we create a special implementation for the case $ A^+ \times B \langle c \rangle$ that avoids computing the full closure $A^+$, as a kind of restricted positive closure that starts instead with $S \gets A \times B \langle c \rangle$. To handle $A^* \times B \langle c \rangle$ we start with $S \gets (E+B)\langle c \rangle$. The cases $\langle r \rangle A \times B^{*/+}$ are handled analogously, as well as the cases with both restrictions. The parser is enhanced to detect those cases. 

Figure~\ref{fig:query-plan} illustrates two relevant cases.

\section{A Baseline} \label{sec:baseline}

We could not find an established software for computations with sparse Boolean matrices, for example to implement transitive closures.
We then implemented a baseline representation of sparse matrices, which combines (and adapts to the Boolean case) the well-known CSR and CSC formats \cite[Sec.~3.4]{Saa03} in order to speed up multiplications. We store a vector of nonempty row numbers and a similar vector of their starting positions in a third, larger, vector. This third vector stores, for each nonempty row, the increasing sequence of the columns of its nonempty cells. Similar (redundant) vectors are stored for the column-wise view.

\begin{figure*}[t]
\centering
\includegraphics[width=0.8\textwidth]{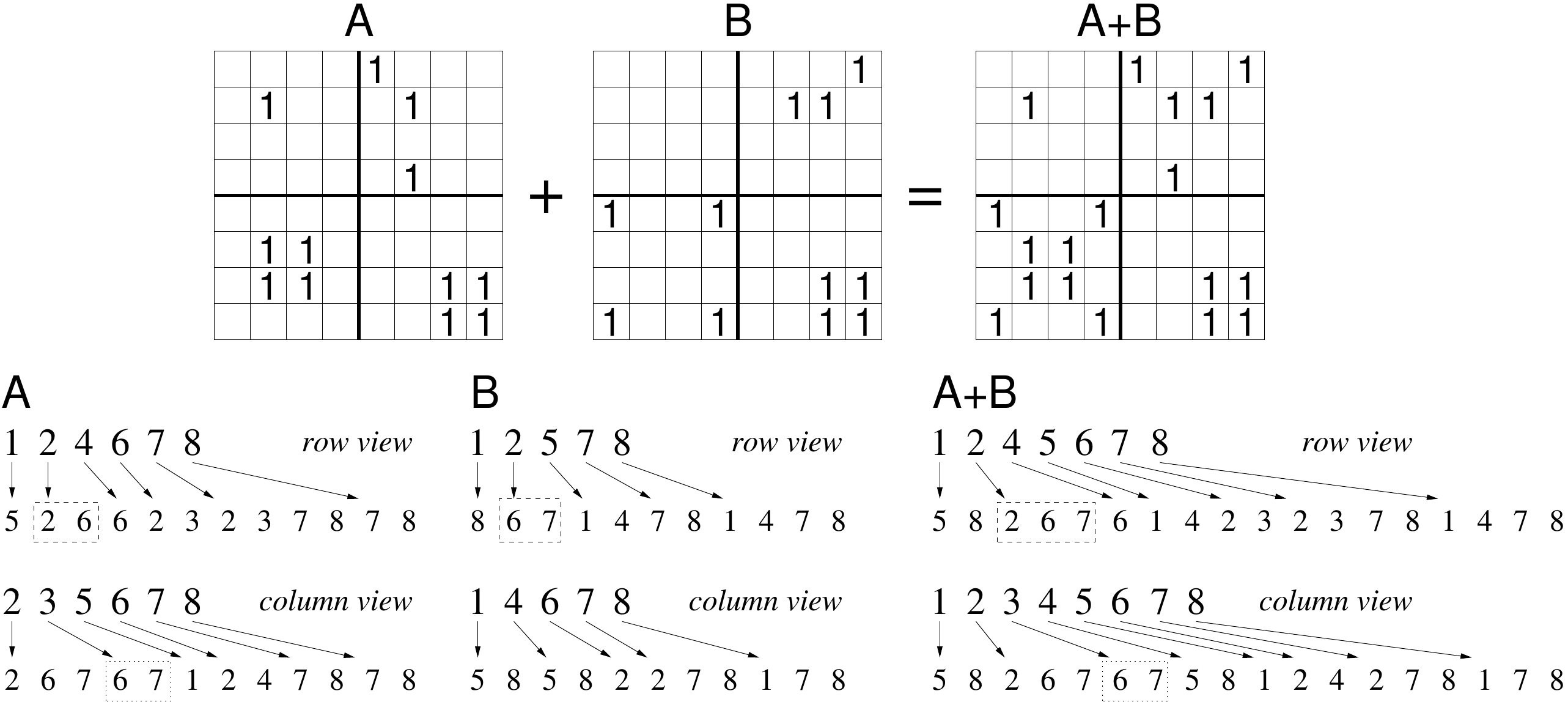}
\caption{The baseline format and summation algorithm. In both the row and column views, we traverse both sets of rows/columns, copy the unique rows/columns (as the dotted box corresponding to column 3 in $A$, whose rows 6 and 7 are copied to the output) and merge the repeated ones (as the dashed boxes corresponding to row 2, whose sets of columns, $\{2,6\}$ and $\{6,7\}$, are merged in the output).}
\label{fig:sumbase}
\end{figure*}

Transpositions are carried out in $O(1)$ time by just exchanging the row-view and the column-view vectors. The Boolean sum $A + B$ merges the nonempty rows, and when the same row appears in both matrices it merges their nonempty columns. The column-view is computed analogously, thus the sum takes time $O(a+b)$. The algorithm is also cache-friendly, as it makes a single left-to-right pass over the input and output arrays. Further, it uses native memory-copy operations when copying whole rows/columns that occur in only one of the matrices, which is faster than merging (despite both operations being linear in the output size). 
Figure~\ref{fig:sumbase} illustrates the sum operation.

\begin{figure*}[t]
\centering
\includegraphics[width=0.8\textwidth]{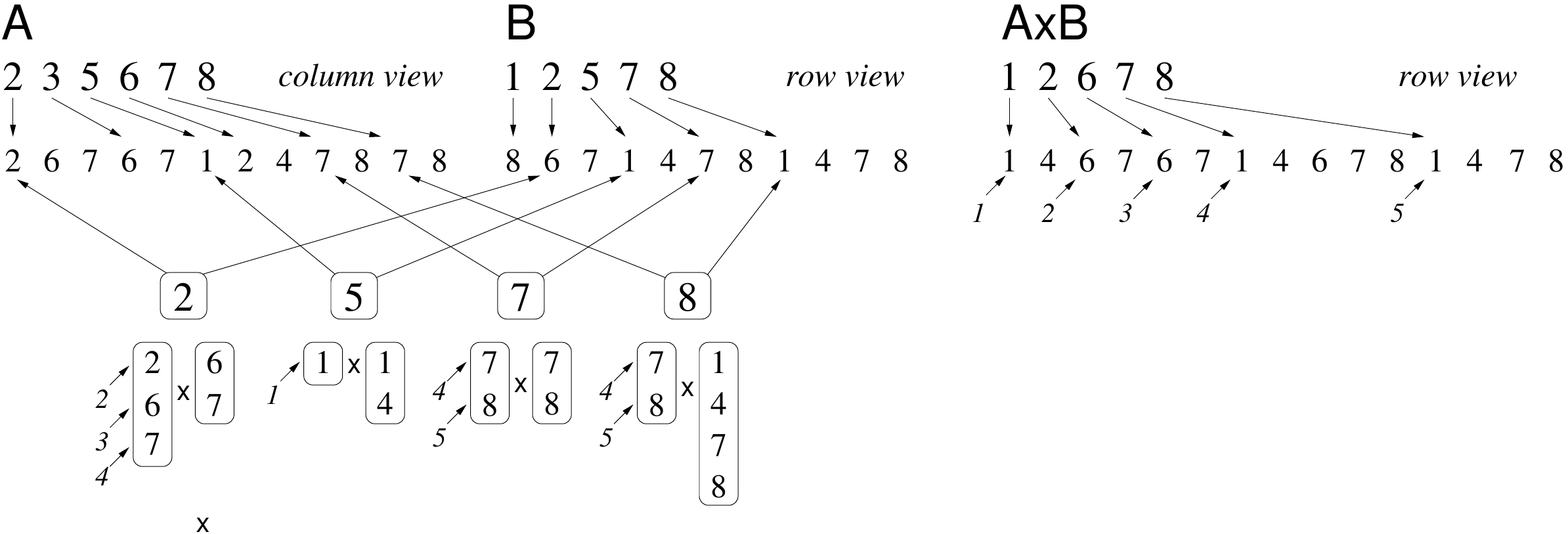}
\caption{Our implementation of Shoor's multiplication algorithm. The row view of $A \times B$ is produced from the column view of $A$ and the row view of $B$. On the bottom-left, the priority queue of 4 tasks created from the intersecting elements $c_i=r_j \in \{2,5,7,8\}$, for each of which we store two pointers towards lists of rows in $A$ and of columns in $B$. Their cartesian products are indicated explicitly on the bottom for illustrative purposes. The diagonal arrows show the 5 steps along which we create the 5 rows of the results, by choosing the smallest remaining row in each task. We first create row $1$, with columns $\{1,4\}$, then row $2$ with columns $\{6,7\}$, then row $6$ with colums $\{6,7\}$. The next row, $7$, appears in tasks $2$, $7$, and $8$, so we merge their sets of columns, $\{6,7\}$, $\{7,8\}$, and $\{1,4,7,8\}$. The last row, $7$, also requires merging from tasks $7$ and $8$.}
\label{fig:multbase}
\end{figure*}

For the Boolean multiplication $A \times B$, we use Schoor's algorithm \cite{Schoor82}, whose average time is $O(ab/v)$ if the 1s are uniformly distributed. Shoor's algorithm intersects the nonempty columns $c_i$ of $A$ with the nonempty rows $r_j$ of $B$. For each pair $c_i=r_j$, it creates the Cartesian product of all the rows associated with $c_i$ in $A$ with all the columns associated with $r_j$ in $B$. The result is the union of the pairs in all those Cartesian products. Our implementation, which is more space-efficient, takes $O(ab\log(v)/v)$ time: we first create the row-wise view of the matrix and at the end use it to generate the column-wise view. To create the row-wise view, we set up a priority queue of {\em tasks} $c_i=r_j$, pointing to the associated rows and columns of $A$ and $B$ and sorted by the smallest row associated with $c_i$. Once the set of all tasks is created, we extract the smallest row from the queue and append to the result all the columns associated with $r_j$ in $B$---we may have to merge several column sequences if they are paired with the same minimum row value, and use another priority queue for that. The use of priority queues yields the $O(\log v)$ additional term in the time complexity. Figure~\ref{fig:multbase} illustrates the algorithm.

% El algoritmo de la verguenza:
%we consider he nonempty columns of every nonempty row of $A$ and the nonempty rows of every nonempty column of $B$, determining if their intersection is empty or not. This is done in a merge-like fashion if both sequences are about the same size, or searching the longer list for the elements of the shorter with exponential search otherwise. For an analysis, consider the average case over uniformly distributed 1s, and assume $a,b \ge v$. Then, for each row and column, we merge $a/v$ with $b/v$ points, in $O(v\min(a,b)\log\frac{\max(a,b)}{\min(a,b)})$ average time. The same complexity is obtained if $\min(a,b)<v<\max(a,b)$; if both are less than $v$, the time is $O(ab)$.
%If $A$ ($B$) has $a_r$ nonempty rows ($b_c$ nonempty columns) and each has $a/a_r$ ($b/b_c$) nonempty cells, then the time is $O(a_r b_c \min(a/a_r,b/b_c)\log\frac{\max(a/a_r,b/b_c)}{\min(a/a_r,b/b_c)})$,
%% = O(\min(a\,b_c,b\,a_r)\log...
%which reaches $O(ab)$ if $a_r=a$ and $b_r=b$.

Closures can be computed naively using $O(\log v)$ multiplications. We implement instead an advanced closure algorithm \cite{Purdom70}, which first computes the strongly connected components (scc) of the graph using Tarjan's algorithm \cite{Tar72}, then creates the reduced and acyclic graph of the scc, computes reachability on the reduced graph in topological order, and finally expands the scc to their node sets. Implemented with the aim of using little working space, the whole algorithm takes time $O(|A^+|\log v)$.

Row and/or column restrictions are handled by restricting the above algorithms to the given row/column; note that finding the desired rows/columns takes just $O(\log v)$ time with the baseline format. Restricted closure operations are performed as for the $k^2$-tree based representation. The parser and its optimizations are also exactly the same.

\section{Experimental Results}

We implemented our scheme in C/C++11 and ran our experiments on an Intel(R) Xeon(R) CPU E5-2630 at 2.30GHz, with 6 cores containing 24 processors in total, 15 MB of cache, and 384 GB of RAM. We compiled using \texttt{g++} with flags \texttt{-std=c++11}, \texttt{-O3}, and \texttt{-msse4.2}. We measure elapsed times.

Our code is publicly available at \texttt{https:// github.com/adriangbrandon/rpq-matrix}.
%\textcolor{red}{indicar donde queda el codigo, en github} \textcolor{brown}{trabajando en esto, esta va a ser la url, pero aun falta comprobar que despues de la limpieza todo funcione: https://github.com/adriangbrandon/rpq-matrix}

%Our basic Baseline implementation handles matrices with up to $2^{32}$ points, using 32-bit integers to address inside the arrays. We also implemented a more general version, called Baseline64, which uses 64-bit integers for addressing them, yet it restricts the cells to the exact number of bits needed. The result is slightly smaller but slower. While Baseline64 matches the functionality of our $k^2$-tree structure (which can handle more than $2^{32}$ points), in the experiments we will use the simple and more efficient Baseline to draw conclusions.

We first study the performance of the individual matrix operations, under various densities, for all our implementations. We then test our implementations in a real scenario where RPQs are solved.

\begin{figure*}[t]
    \centering
    \includegraphics[width=\textwidth]{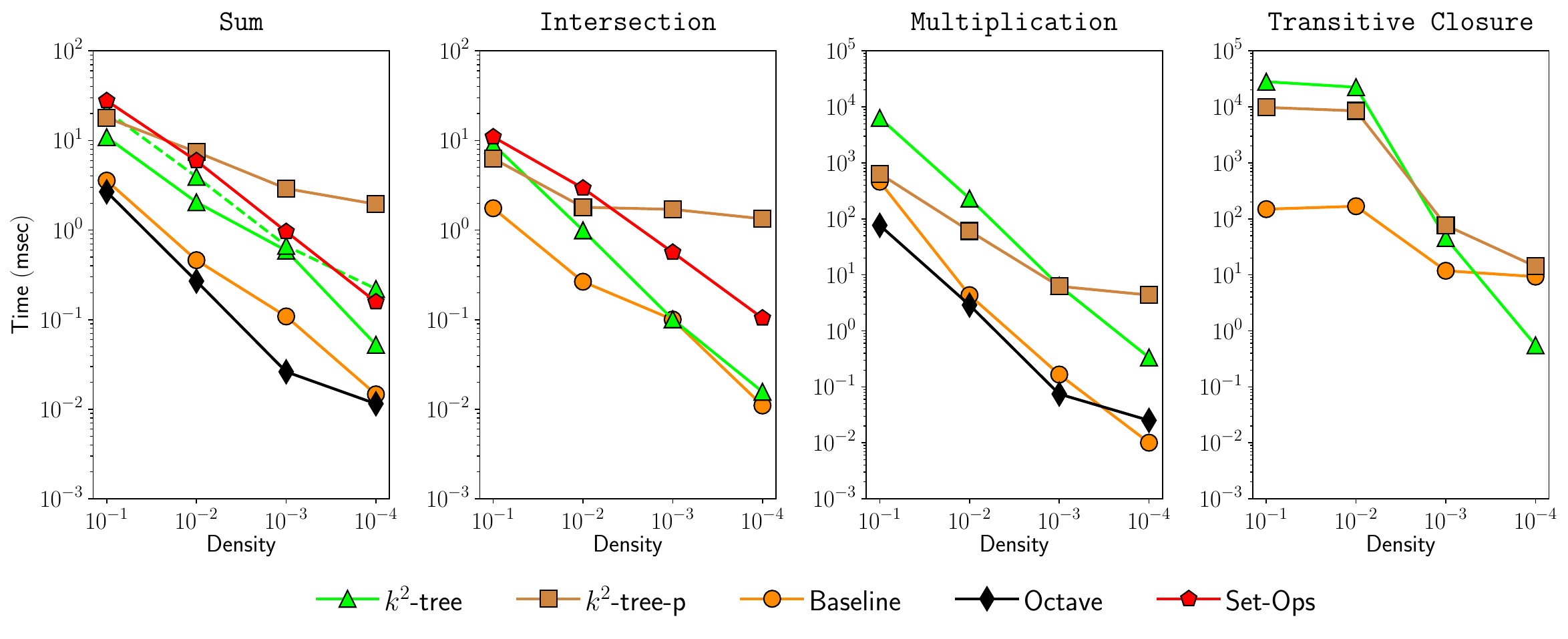}

    \vspace{3mm}
    
    \caption{Averaged times of $k^2$-tree versions and baselines on matrices with different operations and decreasing density. The dashed line denotes the recursive algorithm of the sum in the $k^2$-tree.}
    \label{fig:exp-matrix}
\end{figure*}

\begin{figure*}[t!]
    \centering
    \includegraphics[width=0.8\textwidth]{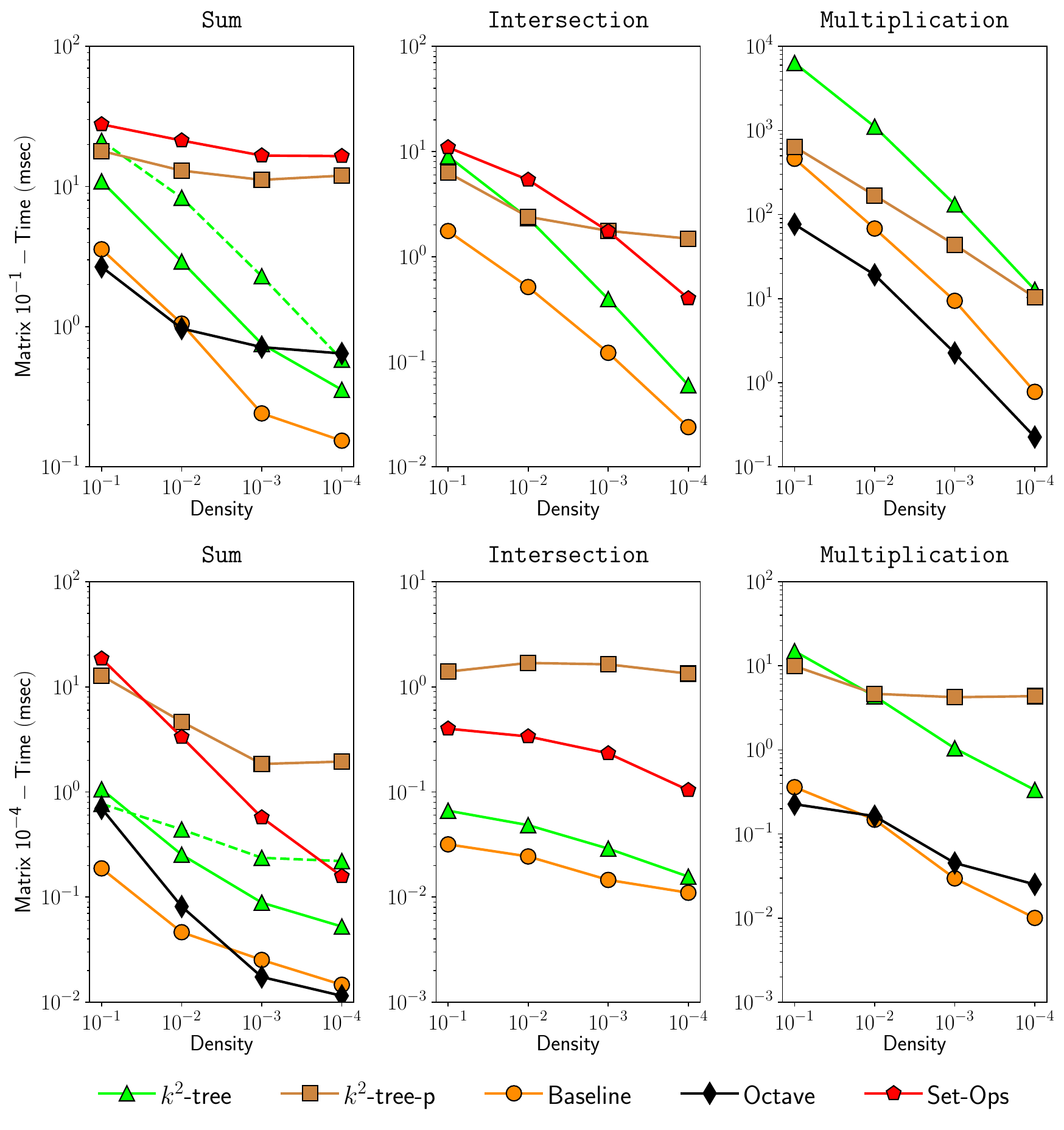}

    \vspace{3mm}
    
    \caption{Average times of $k^2$-tree versions and baselines on matrices with different densities: operated with a matrix of density $10^{-1}$ on top and of density $10^{-4}$ on the bottom. The dashed line denotes the recursive algorithm of the sum in the $k^2$-tree.}
    \label{fig:exp-matrix2}
\end{figure*}

\subsection{Performance of Matrix Operations}
\no{
\begin{figure*}[t]
    \centering
    \includegraphics[width=0.95\textwidth]{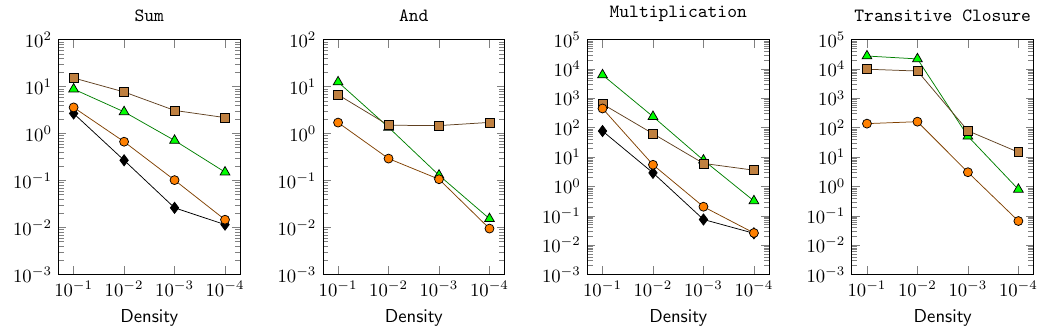}
    \vspace{0.3cm}
       \begin{tikzpicture}
       \begin{customlegend}[legend columns=4,legend style={draw=none,column sep=1ex, font=\footnotesize, cells={anchor=west}, at={(0.5,-0.15)},
      anchor=north, /tikz/every even column/.append style={column sep=0.4cm}},legend entries={\small\textsf{$k^2$-tree},\small\textsf{$k^2$-tree-p},\small\textsf{Baseline},\small\textsf{Octave}}]
    \addlegendimage{mark=triangle*,draw=green!50!black,mark options={
                    draw=black,
                    fill=green,
                }, mark size=3.2pt}
         \addlegendimage{mark=square*, black, draw=black,mark options={
                    draw=black,
                    fill=brown,
                }, mark size=2.6pt}
    \addlegendimage{mark=*,orange!50!black,mark options={
                    draw=black,
                    fill=orange,
                }, mark size=2.6pt}
                \addlegendimage{mark=diamond*,black,mark options={
                    draw=black,
                    fill=black,
                }, mark size=3.4pt}
       \end{customlegend}
       
    \end{tikzpicture}
    \caption{Averaged times of both $k^2$-tree versions and our baseline on matrices with different densities and operations. The dashed line denotes the recursive algorithm of the sum in the $k^2$-tree.}
    \label{fig:exp-matrix}
\end{figure*}}

In order to evaluate the performance of matrix operations, we created $80$ squared matrices with $v$ set to $1{,}000$, where the 1s are uniformly distributed according to different densities: $10^{-1}$, $10^{-2}$, $10^{-3}$, and $10^{-4}$. Specifically, we generated $20$ matrices for each density. We built those matrices with the $k^2$-tree and our baseline. On those systems, for each consecutive pair of matrices of the same density, we ran the operations sum, intersection, and multiplication without any kind of restriction. In addition, the transitive closure is computed on each matrix. Since we have sequential and parallel algorithms for each operation, we denote them as $k^2$-tree and $k^2$-tree-p, respectively. The averaged times of each type operation, separated by density, are shown in Figure~\ref{fig:exp-matrix}. In Figure~\ref{fig:exp-matrix2}, instead, we operate matrices of different densities against a dense matrix (on top) and against a sparse matrix (on the bottom).

We test the sum (Boolean ``or'') and intersection (Boolean ``and'') operations as representatives of the other similar operations. For those, we include in the comparison the existing work that supports set operations \cite{QFPLG19} (called here Set-Ops), which also uses $k^2$-trees to represent the data. Supporting matrix multiplications and transitive closures on $k^2$-trees, instead, is a novelty of our approach. We have not found any other software to compare with, in particular a non-compressed representation of sparse Boolean matrices; this is why we created our baseline, which is also included in the comparison. As a sanity check, we introduce in the comparison a mature software for numeric (not Boolean) computations on sparse matrices: Octave~\cite{octave}, an open-source alternative to MatLab. Octave is compared only for sums and multiplications, whose complexities are similar for numeric versus Boolean matrices. No equivalent to intersections and transitive closures are supported in Octave.

\paragraph{$K^2$-trees.} Figure~\ref{fig:exp-matrix} shows that our sequential algorithm described in Section~\ref{sec:sum} is consistently faster than the recursive one, by a factor of 1.1--4.2, and than the sequential algorithm implemented in Set-Ops, by a factor around 1.6--3.0. The former difference owes to the fact that the sequential algorithm is simpler than the recursive one; the latter difference owes to our improvement when copying submatrices by whole computer words. This is evident in Figure~\ref{fig:exp-matrix2}, where on top the size of the output is still large (a dense matrix), but its alternation measure $\delta$ decreases as the other matrix becomes sparser (recall our fine-grained analysis in Section~\ref{sec:sum}). Both our sequential and recursive algorithms are adaptive to $\delta$ (the recursive algorithm even in terms of complexity) and consequently their times decrease with the size of the sparser matrix; Set-Ops, instead, stays proportional to the output size and independent of $\delta$. On the bottom plots, the output size also decreases (as $\delta$ does) with the size of the denser matrix, and thus all the times decrease accordingly. Here the adaptiveness of our recursive algorithm makes it outperform our sequential algorithm when the difference in densities is maximal, yet its more complex nature makes it finally yield to Set-Ops' sequential algorithm as the difference in densities decreases.

For the other sum-like operations, like the intersection, only a recursive version exists, as explained, both for our algorithm and for Set-Ops'. As expected from Table~\ref{tab:times}, the times are proportional to the number $m$ of 1s in the matrices (i.e., to the densities).  This is explained again by our fined-grained analysis in Section~\ref{sec:sum}, which shows a dependence on the output size more than on the input size (in the case of the sum, both input and output sizes are of the same order); this can also be seen in Figure~\ref{fig:exp-matrix2}. Despite implementing the same algorithm, our code is faster than Set-Ops', with the difference broadening as densities decrease.

Multiplications are in Figure~\ref{fig:exp-matrix} around 3 orders of magnitude slower than sums on the denser matrices, which corresponds to comparing time complexities $m^{3/2}$ with $m$ (see Table~\ref{tab:times}). As $m$ decreases on the sparser matrices, the gap shrinks to one order of magnitude. Figure~\ref{fig:exp-matrix2} shows a notable match with our time complexity analysis of Section~\ref{sec:mult-anal}, $O(\min(a,b)\sqrt{\max(a,b)}\log v)$: the slope on the top plot, where the $x$ axis is the size of $\min(a,b)$, is about twice that of the bottom plot, where the $x$ axis is the size of $\max(a,b)$ (the plots use logarithmic scale).

Finally, transitive closures are about an order of magnitude slower than multiplications, as their time complexity is proportional to $(m^+)^{3/2}$, where $m^+$ is the size of the output. This output size reaches a saturation point at density $10^{-2}$, so the time for density $10^{-1}$ is not very different. The expected density of the transitive closure of a random matrix with density $d > 1/v$ converges to around a constant close to 1 \cite{Kar90}, that is, $m^+$ approaches $v^2$. This is indeed the case of our densities $10^{-1}$ and $10^{-2}$, but not of the sparser matrices.

\paragraph{Parallel $k^2$-trees.}
Table~\ref{tab:times} predicts perfect speedups for $k^2$-tree-p, except for two overheads: one is proportional to the output size and the other is a polylog that grows for the transitive closure operation. In the case of the sum-like operations, the impact of the first overhead is comparable to the parallel time; in addition,  the parallel algorithm must be recursive and cannot write the results directly to the output (recall Section~\ref{sec:parsum}). This combination makes parallelism achieve in Figure~\ref{fig:exp-matrix} only a moderate speedup of 1.8 over Set-Ops on dense matrices, and to always lose to both implementations of the sequential $k^2$-tree. The speedup vanishes, and parallelism becomes actually counterproductive, as densities decrease. This is related to an effect that does not show up in the PRAM analysis: the system overhead incurred when creating the threads reduces the impact of parallelism on the easier operations. The case of intersections is similar, though the $k^2$-tree-p outperforms the $k^2$-tree (by a very slight margin) on the densest matrices. The overhead incurred by parallelization is more visible in the case of intersections, where the resulting matrices are extremely sparse but $k^2$-tree-p is unable to reduce the times obtained for density $10^{-2}$. Figure~\ref{fig:exp-matrix2} also shows that the $k^2$-tree-p cannot exploit the extremely low densities of the output.

On multiplications, the PRAM overheads are less significant because they are of lower order than the amount of work to do---$O(m^2/v)$ versus $O(m^{3/2})$. Figure~\ref{fig:exp-matrix} shows that, on the denser matrices, the speedup is near 10---making the parallel $k^2$-tree approach our Baseline---, but it decreases up to becoming counterproductive on densities below $10^{-3}$. Figure~\ref{fig:exp-matrix2} confirms that the speedup is better when the density of the resulting matrix---which is $10^3 d_1 d_2$ on matrices with densities $d_1$ and $d_2$---is higher.

On transitive closures, perhaps due to the higher polylog overhead, the best speedup obtained by the $k^2$-tree-p is around 3 and decreases as for multiplications.

\paragraph{Baselines.}
Octave is 1.3--3.9 times faster than our baseline on sums. On multiplications, it is up to 6 times faster, but the gap decreases with lower densities, where the baseline finally catches up. In general, we can see that our baseline implementation achieves a reasonably competitive performance against much more mature implementations of sparse matrix operations, thereby providing a relevant implementation of Boolean operations on sparse matrices. Further, the top of Figure~\ref{fig:exp-matrix2} shows that, on sums, Octave is not adaptive to the alternation complexity $\delta$ (which decreases with the size of the sparser matrix), but is instead proportional to the size of the input or the output. As a consequence, our Baseline does outperform Octave---by a margin of up to 4.2---when the densities are very different.\footnote{Further, Octave's times may differ significantly depending on the order in which the two matrices are multiplied ($A\times B$ or $B \times A$), even when they are random and the result has about the same cardinality. We chose the best of both times in the plots (although in real cases one cannot choose).}

In all the operations, our baseline is considerably faster than the $k^2$-tree, as we can expect from its better time complexities in Table~\ref{tab:times}.
For the sum-like operations, where the gap is just $O(\log v)$, the baseline is about 4--9 times faster than the recursive sum and the intersection (they get close in the intersection when there are very few points), and even 3--5 times faster than our sequential implementation of the sum. Figure~\ref{fig:exp-matrix2} shows that the Baseline is also adaptive to the lower densities, just as the $k^2$-tree, because it also processes faster the submatrices that must be copied directly to the output; recall Section~\ref{sec:baseline}.

On the heavier operations, where the time complexities are farther apart (recall that $m^2/v$ is always less than $m^{3/2}$), the baseline is 14--53 times faster for multiplications and 4--189 times faster for transitive closures. An exception is the lowest density, where the recursive algorithm we designed for the $k^2$-tree outperforms the algorithm based in strongly connected components we implemented for the Baseline. This latter algorithm has some basic setup costs that possibly offset its benefits when there are very few 1s in the matrix.

As shown in the next section, the baseline uses in exchange about 4 times more space than the $k^2$-trees. The next section also shows that the large differences exposed here shrink considerably on the real-life application. The reason is that those matrices are far larger and sparser than those we tried here, and as we have seen the differences shrink as the density decreases.

\subsection{Performance on Real-Life RPQs}

We now use our Boolean sparse matrix algebra implementations to solve actual RPQs, from a query log posed to real-world graph database. 
We used a  Wikidata graph~\cite{VrandecicK14} of $n = 958{,}844{,}164$ edges, $v = 348{,}945{,}080$ nodes, 
%$|S| = 106{,}736{,}662$ subjects, 
and $5{,}419$ predicates.
%, and $|O| = 295{,}611{,}216$ objects.
Separating the edges by predicate and 
representing the two nodes of each edge as 32-bit integers, the data set requires 8.5 GB.
% 4*(958844164*2+348945080) bytes
%The space required for this dataset is 10.7~GB in plain form and 7.9 GB in packed form.
%
We compared our implementations with the following  systems:

\begin{itemize}
\item \textit{Ring}: A compact data structure 
that supports RPQs in labeled graphs~\cite{AHNRicde22,AGHNRvldbj24}. The variant {\em Ring$_\text{AB}$} uses more space but is much faster.
\item \textit{Jena}: A reference implementation of the SPARQL standard. 
\item \textit{Virtuoso}: A popular graph database that hosts the public DBpedia endpoint, among others~\cite{Virtuoso}.
\item \textit{Blazegraph}: The graph database system~\cite{Blazegraph} hosting the official Wikidata Query Service~\cite{MalyshevKGGB18}.
\end{itemize}

\begin{table*}[t]
\centering
\caption{Index space (in bytes per triple), indexing time (in hours), and some statistics on the query times (in seconds). Row ``Timeouts'' counts queries that take over 60 seconds or are rejected by the planner as too costly. 2RPQs with some constant node are indicated by $\mathtt{c}$, and without by $\neg\mathtt{c}$.}
\label{tab:statistics}
%\small %OJO: Cambie esto de \small a \tiny 
\setlength{\tabcolsep}{5pt}
\begin{tabular}{lrrrrrrrr}
\toprule
  & $k^2$-tree & $k^2$-tree-p & Baseline %& Baseline64 
           & Ring & Ring\textsubscript{AB} & Jena & Virtuoso & Blazegraph \\
\midrule

Index space (bpt) & 4.33 & 4.33 & 16.45 %& 15.44 
            & 16.41 & 27.93 & 95.83 & 60.07 & 90.79 \\ 
Indexing time (hs) & 0.3 & 0.3 & 5.5 %& 5.5 
            & 7.5 & 8.3 & 37.4 & 3.0 & 39.4 \\ % diccionario 5.22
\midrule
Average (sec) & 3.25 & 3.47 & 1.39 %& 2.03 
            & 1.19 & 0.41 & 4.51 & 2.08 & 3.23 \\
Median (sec) & 0.35 & 0.40 & 0.005 %& 0.005 
            & 0.09 & 0.03 & 0.21 & 0.13 & 0.13 \\
Timeouts & 39 & 44 & 14 %& 17 
            & 9 & 1 & 84 & 1 & 41 \\
\midrule
Average $\mathtt{c}$ (sec) & 2.84 & 3.10 & 1.19 %& 1.59 
            & 0.65 & 0.25 & 3.62 & 1.79 & 3.24 \\
Median $\mathtt{c}$ (sec) & 0.35 & 0.40 & 0.005 %& 0.005 
            & 0.08 & 0.03 & 0.19 & 0.11 & 0.13 \\
Timeouts $\mathtt{c}$ & 30 & 33  & 12 %& 14 
            & 2 & 0 & 58 & 1 & 39 \\
  \midrule
Average $\neg\mathtt{c}$ (sec) & 11.92  & 11.21 & 5.45 %& 10.32 
            & 12.43 & 3.66 & 22.83 & 8.17 & 2.98\\
Median $\neg\mathtt{c}$ (sec) & 0.87 & 0.79 & 0.01 %& 0.03 
            & 2.09 & 0.93 & 1.57 & 3.89 & 0.14\\
Timeouts $\neg\mathtt{c}$ & 9 & 11 & 2 %& 3 
            & 7 & 1 & 26 & 0 & 6  \\
\bottomrule
\end{tabular}
%\vspace{0.2cm}
%\vspace{-5mm}
\end{table*}

To evaluate complex real-world 2RPQs, we extracted all 2RPQs that were not simple labels, from the code-500 (timeout) sections of the seven intervals of the Wikidata Query Logs~\cite{MalyshevKGGB18}. 
We then normalized variable names and removed disrupting queries: duplicated queries and queries producing more than $10^6$ results for compatibility with Virtuoso. The result was $1{,}567$ unique queries.

We ran the queries in each system with a timeout limit of 60 seconds. On the $k^2$-tree representation, we ran the single- and multi-thread versions of our algorithms. Table~\ref{tab:statistics} summarizes the space usage and time performance of all the systems. Notably, our $k^2$-tree based approach yields the most compact structure, requiring only 4.33 bytes per triple (bpt). This is nearly half the space of the described plain representation of the raw data, and about a fourth of the space used by the next smallest representations that support 2RPQs (Ring and our Baseline). Classical systems use 14--22 times more space than our $k^2$-trees. Note also that the $k^2$-tree representation is 1--2 orders of magnitude faster to build than the others.

This reduced space is paid in terms of time performance. Our sequential structure is on average 2.7 times slower than the Ring, 7.9 times slower than the fastest system (Ring$_\text{AB}$), and 1.6 times slower than the fastest classical system (Virtuoso). Ring$_\text{AB}$ and Virtuoso stand out for their stability---just one timeout. Still, the $k^2$-tree solves those 2RPQs in less than 4 seconds on average, and are competitive with established systems like Blazegraph and Jena. The median $k^2$-tree time is 1.7--11.7 times higher than the others. Our $k^2$-tree-p does not compete in general with the sequential $k^2$-tree, which is to be expected in principle because the matrices are very sparse in this application.

Our Baseline, on the other hand, uses almost the same space as the Ring, and it is on average 16\% slower. It is 2.3 times faster than $k^2$-trees, and 3.4 times slower than the Ring$_\text{AB}$ (which uses 1.7 times more space, however). While using 3.7 less space than Virtuoso, it is 49\% faster. While the Baseline shows no advantage over the Ring in those aspects, it solves many easy queries much faster than all the systems---its median is an order of magnitude lower. Yet, it produces more timeouts than the Ring, Ring$_\text{AB}$, and Virtuoso, thereby displaying less stability. The other systems time out on many more queries, though. 

The situation turns more against our matrix-based methods on the easier 2RPQs---those containing some constant. On those, the single-threaded $k^2$-tree is on average 4.4 times slower than the Ring, 12.4 times slower than the Ring$_\text{AB}$, and 1.6 times slower than Virtuoso, yet still outperforming Blazegraph and Jena. The Baseline is here 1.8 times slower than the Ring and 4.8 times slower than the Ring$_\text{AB}$, yet it is still 1.5 times faster than Virtuoso, and it is still an order of magnitude faster on the median.

On the harder queries, with no constant extreme, the relative performance of matrix-based methods is much better: the $k^2$-tree and $k^2$-tree-p are 5\% and 11\% faster than the Ring, respectively, still using 4 times less space. This time the parallel version reduces the median times, as most queries are hard enough to benefit from parallelism. The Baseline, still using about the same space as the Ring, is 2.3 times faster. In these queries, however, the fastest system is Blazegraph, which is 1.8 faster than our Baseline and 3.8 times faster than $k^2$-tree-p. The Ring$_\text{AB}$, using not as much space, is 1.5 times faster than our Baseline and 3.1 times faster than the $k^2$-tree-p. Yet, this comes at the expense of using 1.7 times more space than the Baseline and 6.5 times more space than $k^2$-trees. The Baseline still has 1--2 orders of magnitude faster median times.

\begin{figure*}[t!]
    \centering
    \includegraphics[width=0.95\textwidth]{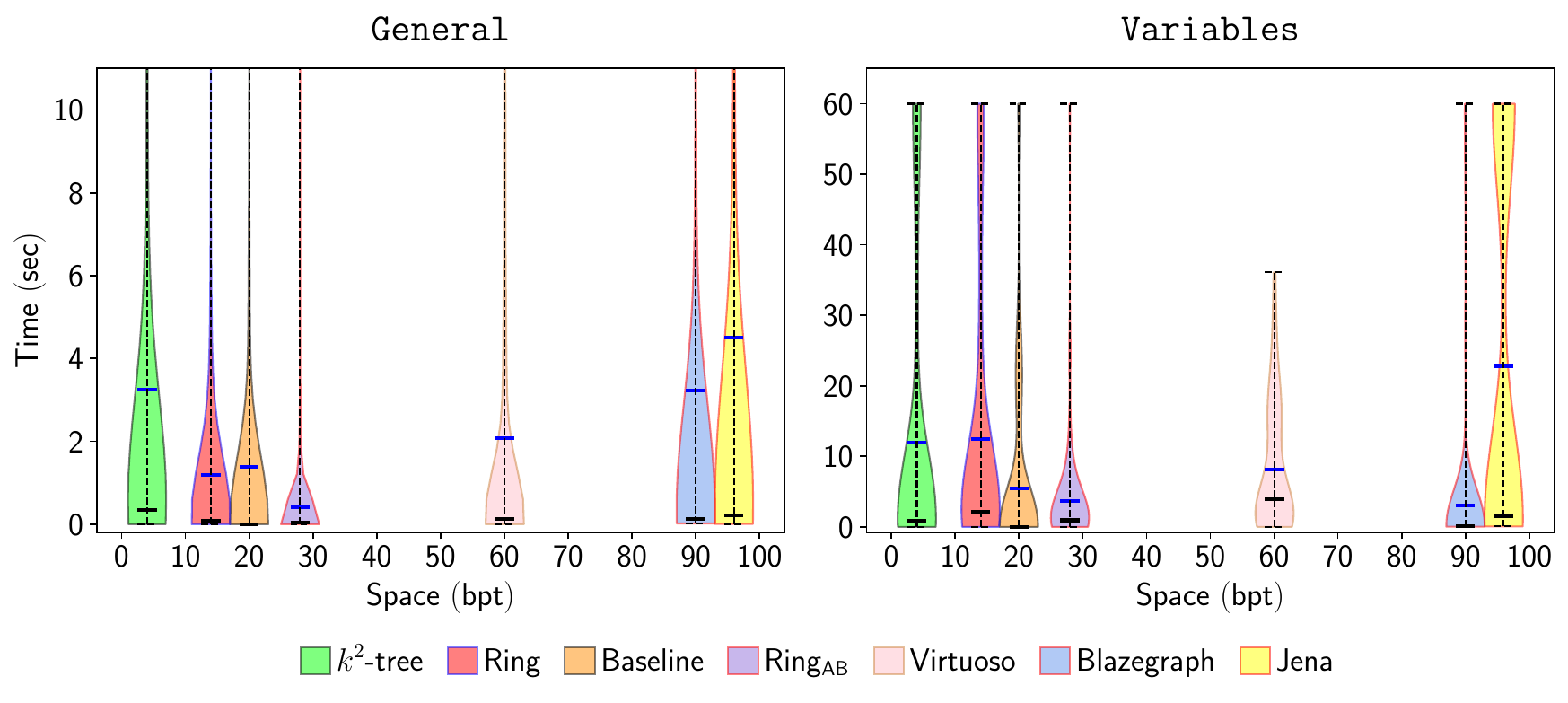}
    \caption{Space and query time distribution of the systems in general (left) and for the 2RPQs with no constants (right). The baseline and the Ring use almost the same space, but we separate them for readability.}
    \label{fig:spacetime}
\end{figure*}

Figure~\ref{fig:spacetime} shows the query time distributions using violin plots \cite{violinplot} (these show a symmetric histogram of values along the $y$ axis) together with averages (higher segments) and medians (lower segments). The violin plots are placed, along the $x$ coordinate, corresponding approximately to the space usage of the different structures---see Table~\ref{tab:statistics} for the detailed space usage. The left plot refers to all the queries, whereas the right one considers only the harder 2RPQs (with both variable extremes). We leave out the $k^2$-tree-p, as it performs almost identically to the $k^2$-tree.

In the general case, the $k^2$-tree and both Ring variants are the dominant representations. The first one offers a decent solution with low space (slightly over 4 bpt), solving 90\% of the queries in less than 5.3 seconds. Using 4 times that space (about 16 bpt), the Ring is noticeably faster than the $k^2$-trees and distributes slightly better than the Baseline, though the latter has a much lower median. Finally, the Ring$_\text{AB}$ uses about twice that space (28 bpt) and is considerably faster and more stable than the Ring. Classical systems are all outperformed by the Ring$_\text{AB}$ in both time and space.

When it comes to handling the harder 2RPQs, the right plot shows that the Baseline becomes noticeably faster than the Ring---which distributes even worse than $k^2$-trees and thus becomes not competitive. The Baseline's distribution is only outperformed---by a discrete margin---by that of Ring$_\text{AB}$ and Blazegraph, yet using much more space (about 28 and 90 bpt, respectively). 

\no{
\begin{figure*}[t]
    \centering
    \includegraphics[width=0.95\textwidth]{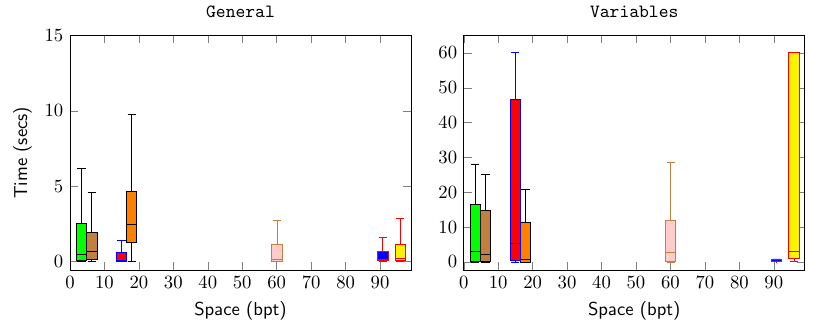}
    \vspace{0.3cm}
       \begin{tikzpicture}
       \begin{customlegend}[legend columns=7,legend style={draw=none,column sep=1ex, font=\footnotesize, cells={anchor=west}, at={(0.5,-0.15)},
      anchor=north, /tikz/every even column/.append style={column sep=0.4cm}},legend entries={\small\textsf{$k^2$-tree},\small\textsf{$k^2$-tree-p},\small\textsf{Baseline},\small\textsf{Ring},\small\textsf{Jena}, \small\textsf{Virtuoso}, 
       \small\textsf{Blazegraph}}]
        \addlegendimage{black,fill=green, ybar stacked}
         \addlegendimage{black,fill=brown, ybar stacked}
        \addlegendimage{black,fill=orange, ybar stacked}
          \addlegendimage{blue,fill=red, ybar stacked}
          \addlegendimage{red,fill=yellow,ybar stacked}
          \addlegendimage{brown,fill=red!20,ybar stacked}
          \addlegendimage{red,fill=blue,ybar stacked}
       \end{customlegend}
    \end{tikzpicture}
    \caption{Space and query time distribution of the systems in general (left) and for the 2RPQs with no constants (right). Both $k^2$-tree versions use the same space; the baseline and the Ring also use almost the same space.\adrian{Falta actualizar}}
    \label{fig:spacetime}
\end{figure*}}

\section{Conclusions}

We have explored the use of a Boolean matrix algebra to implement Regular Path Queries (RPQs) on graph databases. This path is usually disregarded because the matrix sizes are quadratic on the number of graph nodes, but we exploit their sparsity to sidestep this issue. Our experiments show that even our baseline (i.e., uncompressed) sparse matrix representation uses the same space of the most compact among previous representations, and outperforms them on the most difficult RPQs (i.e., those with no constant ends). We also develop a more compressed sparse matrix representation based on $k^2$-trees, which is four times smaller than the baseline and, although slower, it still handles most RPQs within a few seconds. We have implemented and adapted state-of-the-art algorithms for sparse matrices to implement our baseline, and designed new ones for the $k^2$-trees. Our sparse Boolean matrix algebra implementations, which are publicly available, are of interest beyond solving RPQs, as they arise in other situations, such as ML applications \cite{EBHRR19}.

Immediate extensions to our work are the implementation of negated labels, which require a nonexpensive way to represent and handle submatrices full of 1s. Such extensions of $k^2$-trees have been proposed \cite{BABNPspire13,QFPLG19}, but they have not been adapted to handle the most complex Boolean matrix operations (multiplication and transitive closure). We can also strenghten our RPQ optimizer in order to detect common subexpressions and exploit a number of identities of the Boolean algebra we have disregarded for now.
%e.g. $A|A$, $A \times I$, $A+0$, $A \times 0$, $I^*$, and other identities of the Boolean algebra.

Another path of future work is to integrate RPQs with BGPs, the other main segment of most graph query languages. In those combined queries, some triple patterns $(x,p,y)$ refer to predicates $p$ and others may be 2RPQs of the form $(x,E,y)$. We can then use our matrix algebra to materialize those RPQs into a resulting matrix, which acts as a new predicate $p_E$. The result is a simple BGP, which can then be solved with Qdags \cite{ANRRtods22}, an existing solution BGPs that is also based on representing each predicate as a $k^2$-tree. In this way we would not need any extra space, since both indices use exactly the same data structures.

\bibliographystyle{splncs04}
\bibliography{ref-vldb}

\end{document}